\documentclass[12pt,x11names]{article}
\usepackage{latexsym,amssymb,amsmath,amsthm}
\usepackage{graphicx,xcolor,shadow}
\usepackage{color}

\numberwithin{equation}{section}

\def\argmin{\displaystyle\mathop {\mbox{argmin}}}

\def\argmax{\displaystyle\mathop {\mbox{argmax}}}

\newtheorem{assumption}{Assumption}

\newtheorem{defin}{Definition}[section]

\newtheorem{theorem}{Theorem}[section]

\newtheorem{lem}{Lemma}[section]

\newtheorem{corollary}{Corollary}[section]

\theoremstyle{definition}
\newtheorem{example}{Example}[section]
\newtheorem{remark}{Remark}[section]

\newcommand{\CC}{{\mathbb{C}}}

\def\argmin{\displaystyle\mathop {\mbox{\rm argmin}}}

\def\argmax{\displaystyle\mathop {\mbox{\rm argmax}}}

\def\real{\mathbb R}

\def\CC{\mathbb C}

\newcommand{\ba}{{\bf a}}

\newcommand{\bd}{{\bf d}}

\newcommand{\bb}{{\bf b}}
\newcommand{\by}{{\bf y}}

\newcommand{\br}{{\bf r}}
\newcommand{\bc}{{\bf c}}
\newcommand{\be}{{\bf e}}
\newcommand{\bo}{{\bf 0}}
\newcommand{\bx}{{\bf x}}

\newcommand{\bz}{{\bf z}}

\newcommand{\bbi}{{\bf I}}

\newcommand{\bbj}{{{\bf J}}}

\newcommand{\bbq}{{\bf Q}}

\newcommand{\bba}{{\bf A}}

\newcommand{\sgn}{\mbox{sgn\,}}

\title{Sparsity Constrained Nonlinear Optimization:\\Optimality Conditions and Algorithms}
\author{Amir Beck\thanks{Faculty of Industrial Engineering and Management, 
Technion - Israel Institute of Technology, Haifa 3200, Israel. Email: becka@ie.technion.ac.il} \and Yonina C. Eldar\thanks{Faculty of Electrical Engineering, Technion - Israel Institute of Technology, Haifa 32000, Israel. Email: yonina@ee.technion.ac.il}}

\date{\today}
\begin{document}
\maketitle
\begin{abstract} This paper treats the problem of minimizing a general continuously differentiable function subject to sparsity constraints. We present and analyze several different optimality criteria which are based on the notions of stationarity and coordinate-wise optimality. These conditions are then used to derive three numerical algorithms aimed at finding points satisfying the resulting optimality criteria: the iterative hard thresholding method and the greedy and partial sparse-simplex methods. The first algorithm is essentially a gradient projection method while the remaining two algorithms are of coordinate descent type. The theoretical convergence of these methods and their relations to the derived optimality conditions are studied. The algorithms and results are illustrated by several numerical examples.
\end{abstract}
\section{Introduction}

Sparsity has long been exploited in signal processing, applied mathematics, statistics and computer science for tasks such as compression, denoising, model selection, image processing and more \cite{D98,D95,GR97,M08,OF96,JPEG2000,T96}.
Recent years have witnessed a growing interest in sparsity-based processing methods and algorithms for sparse recovery \cite{BF12,BT09-book,TW10}. Despite the great interest in exploiting sparsity in various applications, most of the work to date has focused on recovering sparse data represented by a vector $\bx \
\in \real^n$ from linear measurements of the form $\bb=\bba\bx$. For example, the rapidly growing field of compressed sensing \cite{D06,CRT06,DDEK11} considers recovery of a sparse $\bx$ from a small set of linear measurements $\bb \in \real^m$ where $m$ is usually much smaller than $n$.
Since in practice the measurements are contaminated by noise, a typical approach to recover $\bx$ is to seek a sparse vector $\bx$ that minimizes the quadratic function $\|\bba\bx-\bb\|_2^2$.

In this paper we study the more general problem of minimizing a continuously differentiable objective function subject to a sparsity constraint. More specifically, we consider the problem
$$ \mbox{(P):} \quad \begin{array}{ll} \min & f(\bx) \\ \mbox{s.t.} & \|\bx\|_0 \leq s, \end{array}$$
where $f:\real^n \rightarrow \real$ is a continuously differentiable function, $s>0$ is an integer smaller than $n$ and $\|\bx\|_0$ is the $\ell_0$ norm of $\bx$, which counts the
 number of nonzero components in $\bx$. We do not assume that $f$ is a convex function.
  This, together with the fact that the constraint function is nonconvex, and is in fact not even continuous, renders the problem quite difficult.
Our goal in this paper is to study necessary optimality conditions for problem (P) and to develop algorithms that find points satisying these conditions for general choices of $f$.

Two instances of problem (P) that have been considered in previous literature and will serve as prototype models throughout the paper are described in the following two examples.
\begin{example}[Compressive Sensing] As mentioned above, compressed sensing is concerned with recovery of a sparse vector $\bx$ from linear measurements
$ \bba \bx = \bb$, where $\bba \in \real^{m \times n}, \bb \in \real^m$ and $m$ is usually much smaller than $n$. It is well known that under suitable
conditions on $\bba$, only the order of $s \log n$ measurements are needed to recover $\bx$ \cite{V11}. When noise is present in the measurements, it is natural to consider the corresponding optimization problem (P)
with the objective function given by
$$ f_{\rm LI}(\bx) \equiv \|\bba \bx - \bb\|^2.$$
A variety of algorithms have been proposed in order to approximate the solution to this problem \cite{T04,TW10}. One popular approach is to replace the $\ell_0$ norm with the convex $\ell_1$ norm, which results in a convex problem. A variety of different greedy methods have also been proposed, such as the matching pursuit (MP) and orthogonal MP (OMP) algorithms \cite{MZ93}. We will relate our methods to these approaches in Section~\ref{subsec:mp}. Another method that was proposed recently and is related to our approach below is the iterative hard thresholding  algorithm \cite{BD08}, also referred to as the ``M-sparse" method. In \cite{BD08} the authors consider a majorization-minimization approach to solve (P) with $f=f_{\rm LI}$, and show that the resulting method converges to a local minima of (P) as long as the spectral norm of $\bba$ satisfies $\|\bba\|<1$. This algorithm is essentially a gradient projection method with stepsize 1. In Section~\ref{sec:ssparse} we will revisit the iterative hard thresholding method and show how it can be applied to the general formulation $(P)$, as well as discuss the quality of
the limit points of the sequence generated by the algorithm.
\end{example}

Although linear measurements are the most popular in the literature, recently, attention has been given to quadratic measurements. Sparse recovery problems from quadratic measurements arise in a variety of different problems in optics, as we discuss in the next example.
\begin{example} Recovery of sparse vectors from quadratic measurements has been treated recently
 in the context of sub-wavelength optical imaging \cite{SEA12,SESS11}. In these problems the goal is to recover a sparse image from its far-field measurements, where due to the laws of physics the relationship between the (clean) measurement and the unknown image is quadratic.
  In \cite{SESS11} the quadratic relationship is a result of using partially-incoherent light.
   The quadratic behavior of the measurements in \cite{SEA12} is a result of coherent diffractive imaging in which the image
   is recovered from its intensity pattern.
   Under an appropriate experimental setup, this problem amounts to reconstruction of a sparse signal from the magnitude of its Fourier transform.

 Mathematically, both problems can be described as follows: Given $m$ symmetric matrices $\bba_1,\ldots,\bba_m \in \real^{n \times n}$, find a vector $\bx$ satisfying:
\begin{eqnarray*} \bx^T \bba_i \bx &\approx& c_i, \quad i=1,\ldots,m,\\
\|\bx\|_0  &\leq& s.
\end{eqnarray*}
This problem can be written in the form of problem (P) with
 $$f_{\rm QU}(\bx) \equiv \sum_{i=1}^m \left (\bx^T \bba_i \bx -c_i \right )^2. $$
 In this case, the objective function is nonconvex and quartic.

Quadratic measurements appear more generally in  phase retrieval problems, in which a signal $\bx$ is to
     be recovered from the magnitude of its measurements  $y_i=|\bd_i^* \bx|$, where each measurement is a linear transform of the input $\bx \in \real^n$. Note that $\bd_i$ are complex-valued, that is $\bd_i \in \CC^n$.
Denoting by $b_i$ be the corresponding noisy measurements, and assuming a sparse input, our goal is to minimize
      $\sum_{i=1}^m (b_i^2-|\bd_i^*\bx|^2)^2$
      subject to the constraint that $\| \bx\|_0 \leq s$ for some $s$, where $m$ is the number of measurements. The objective function has the same structure as $f_{\rm QU}$
      with $\bba_i = \Re(\bd_i)\Re(\bd_i)^T+\Im(\bd_i)\Im(\bd_i)^T$.
       In \cite{SESS11}, an algorithm was developed to treat such problems based on a semidefinite relaxation, and low-rank matrix recovery.
        However, for large scale problems, the method is not efficient and difficult to implement.
        An alternative algorithm was designed in \cite{SEA12} based on a greedy search. This approach requires solving a nonconvex optimization program in each internal iteration.

To conclude this example, we note that the problem of recovering a signal from the
magnitude of its Fourier transform has been studied extensively in the literature.
Many methods have been developed for phase recovery \cite{H89} which often rely on  prior information about the signal,
such as positivity or support constraints. One of the most popular techniques is based on alternating projections,
where the current signal estimate is transformed back and forth between the object and the Fourier domains.
The prior information and observations are used in each domain in order to form the next estimate.
 Two of the main approaches of this type are Gerchberg-Saxton \cite{GS72} and Fienup \cite{F82}.
  In general, these methods are not guaranteed to converge, and often require careful parameter selection and sufficient signal
  constraints in order to provide a reasonable result.
\end{example}

In this paper we present a uniform approach to treating problems of the form (P). Necessary optimality conditions for problems consisting of minimizing differentiable (possibly nonconvex) objective functions over convex feasibility sets are well known \cite{B99}. These conditions are also very often the basis for efficient algorithms for solving the respective optimization problems.
However, classical results on nonconvex optimization do not cover the case of sparsity constraints, which are neither convex nor continuous. In Section~\ref{sec:opt} we derive 3 classes of necessary optimality conditions for problem (P): basic feasibility, $L$-stationarity, and coordinate-wise (CW) optimality. We then show that CW-optimality  implies $L$-stationarity for suitable values of $L$, and they both imply the basic feasibility property. In Section~\ref{sec:alg} we  present two classes of algorithms for solving (P). The first algorithm is a generalization of the iterative hard thresholding method, and is based on the notion of $L$-stationarity. Under appropriate conditions we show that the limit points of the method are $L$-stationary points.
The second class of methods are based on the concept of CW-optimality. These are basically coordinate descent type  algorithms which update the support at each iteration by one or two variables.  Due to their resemblance with the celebrated simplex method for linear programming, we refer to these methods as ``sparse-simplex" algorithms. As we show, these algorithms are as simple as the iterative hard thresholding  method, while obtaining stronger optimality guarantees. In Section~\ref{sec:proof} we prove the convergence results of the various algorithms, establishing that the limit points of each of the methods satisfy the respective necessary optimality conditions.

\section{Necessary Optimality Conditions}
\label{sec:opt}

\subsection{Notation and Assumptions}
For a given vector $\bx \in \real^n$ and an index set $R \subseteq \{1,\ldots,n\}$, we denote by $\bx_{R}$ the subvector of $\bx$
corresponding to the indices in $R$. For example, if $\bx = (4,5,2,1)^T$ and $R = \{ 1,3\}$, then $\bx_R = (4,2)^T$. The support set of
$\bx$ is defined by
 $$ I_1(\bx) \equiv \left \{ i : x_i \neq 0 \right \},$$
 \noindent and its complement is
 $$ I_0(\bx)  \equiv \left \{ i: x_i =0 \right \}.$$
We denote by $C_s$ the set of vectors $\bx$ that are at most $s$-sparse:$$C_s=\{\bx : \|\bx\|_0\leq s\}.$$
For a vector $\bx \in \real^n$ and $i \in \{1,2,\ldots,n\}$, the $i$th largest absolute value component in $\bx$ is denoted by $M_i(\bx)$, so that in particular
 $$ M_1(\bx) \geq M_2(\bx) \geq \ldots \geq M_n(\bx).$$
 Also, $M_1(\bx) = \max_{i=1,\ldots,n} |x_i|$ and  $M_n(\bx) = \min_{i=1,\ldots,n} |x_i|.$

Throughout the paper we make the following assumption.
\begin{assumption} \label{as:lb} The objective function $f$ is lower bounded. That is, there exists $\gamma \in \real$ such that $f(\bx) \geq \gamma$ for all $\bx \in \real^n$.
\end{assumption}

\subsection{Basic Feasibility}

Optimality conditions have an important theoretical role in the study of optimization problems. From a practical point of view, they are the basis
for most  numerical solution methods. Therefore, as a first step in studying problem (P), we would like to consider its optimality conditions, and then use them to generate algorithms.
 However, since (P) is nonconvex, it does not seem to posses necessary and sufficient conditions for optimality. Therefore, below we derive several necessary conditions, and analyze the relationship between them. We will then show in Section~\ref{sec:alg} how these conditions lead to algorithms that are guaranteed to generate a point satisfying the respective conditions.

For unconstrained differentiable problems, a necessary optimality condition is that the gradient is zero. It is therefore natural to expect that a similar necessary condition will be true over the support $I_1(\bx^*)$ of an optimal point $\bx^*$. Inspired
by linear programming terminology, we will call a vector satisfying this property a  \textit{basic feasible} vector.

\begin{defin} A vector $\bx^*\in C_s$ is called  a basic feasible (BF) vector of (P) if:
\begin{enumerate}
\item when $\|\bx^*\|_0 <s$, $\nabla f(\bx^*)=0$;
\item when $\|\bx^*\|_0=s$, $ \nabla_i f(\bx^*)=0$ for all $i \in I_1(\bx^*)$.
\end{enumerate}
\end{defin}

We will also say that a vector satisfies the ``basic feasibility property" if it is a BF vector. Theorem~\ref{thm:bfs} establishes the fact that any optimal solution of (P) is also a BF vector.
\begin{theorem} \label{thm:bfs} Let $\bx^*$ be an optimal solution of (P). Then $\bx^*$ is a BF vector. \end{theorem}
\noindent {\bf Proof.} If $\|\bx^*\|_0<s$, then for any $i \in \{1,2,\ldots,n\}$
$$ 0 \in \argmin \{ g(t) \equiv f(\bx^*+t \be_i)\}.$$
\noindent Otherwise there would exist a $t_0$ for which $f(\bx^*+t_0 \be_i)<f(\bx^*)$, which is a contradiction to the optimality of $\bx^*$.
Therefore, we have $\nabla_i f(\bx^*)=g'(0)=0$.
If $\|\bx^*\|_0=s$, then the same argument holds for any $i \in I_1(\bx^*)$. \qed

We conclude that a necessary condition for optimality is basic feasibility. It turns out that this condition is quite weak, namely, there are many BF points that are not optimal points. In the following two subsections we will consider stricter necessary optimality conditions.

Before concluding this section we consider in more detail the special case of $f(\bx)\equiv f_{\rm LI}(\bx) \equiv \| \bba \bx-\bb\|^2$.    We now show that under a suitable condition on $\bba$,
which we refer to as $s$-\textit{regularity},
there are only a finite number of BF points. This implies that there are only a finite number of points suspected to be optimal solutions.
\begin{defin}[$s$-regularity] A matrix $\bba \in \real^{m \times n}$ is called \textit{$s$-regular} if for every index set $I \subseteq \{1,2,\ldots,n\}$ with $|I|=s$, the columns of $\bba$
associated with the index set $I$ are linearly independent.
\end{defin}

\begin{remark}
$s$-regularity can also be expressed in terms of the Kruskal rank of $\bba$: The Kruskal rank of a matrix $\bba$ is equal to $s$ if every $s$ columns of $\bba$ are linearly independent.
Another way to express this property is via the spark -- $\rm{spark}(\bba)$ is the minimum number of linearly dependent columns (see \cite{DE03}). Thus, $\bba$ is $s$-regular if and only if $\rm{spark}(\bba) \geq s+1$.
\end{remark}

When $s \leq m $, the $s$-regularity property is rather mild in the sense that if the components of $\bba$ are independently randomly generated from a continuous distribution,
then the $s$-regularity property will be satisfied with probability one.

It is interesting to note that in the compressed sensing literature, it is typically assumed that $\bba$ is $2s$-regular. This condition is necessary in order to ensure uniqueness of the solution to $\bb=\bba\bx$ for any $\bx$ satisfying $\|\bx\|_0 \leq s$. Here we are only requiring $s$-regularity, which is a milder requirement.

The next lemma shows that when the $s$-regularity property holds, the number of BF points is finite.
\begin{lem} \label{lem:lfinite} Let $f(\bx) \equiv f_{\rm LI}(\bx) = \|\bba \bx-\bb\|^2$, where  $\bba \in \real^{m \times n}$ is an $s$-regular matrix and $\bb \in \real^m$.
Then the number of BF points  of problem (P) is finite.
\end{lem}
\noindent {\bf Proof:} Any BF vector $\bx$ satisfies
$$ \|\bx\|_0 \leq s \mbox{ and }  \nabla_i f_{\rm LI} (\bx) =\bo,\quad i \in I_1(\bx).$$
Denote the support set of $\bx$ by $S = I_1(\bx).$ Then $|S| \leq s$ and from the derivative condition,
$$ \bba_S^T (\bba_S \bx_S-\bb)=\bo,$$
\noindent where $\bba_S$ is the submatrix of $\bba$ comprised of the columns corresponding to the set $S$. Here we used the fact that $\bba\bx=\bba_S\bx_S$ for any $\bx$ with support $S$.
By the $s$-regularity assumption it follows that the matrix
$\bba_S^T \bba_S$ is nonsingular. Thus,
$$\bx_S =  (\bba_S^T \bba_S)^{-1}\bba_S^T \bb.$$
To summarize, for each set of indices $S$ satisfying $|S|\leq s$, there is at most one candidate for a BF vector with support $S$.
 Since the number of
subsets of $\{1,2,\ldots,n\}$ is finite, the result follows. \qed

\subsection{$L$-Stationarity}

As we will see in the examples below, the basic feasibility property is a rather weak necessary optimality condition. Therefore, stronger necessary conditions are needed in order to obtain higher quality solutions.
In this section we consider the $L$-stationarity property which is an extension of the concept of stationarity for convex constrained problems. In the next section we discuss coordinate-wise optimality which leads to stronger optimality results.

We begin by recalling some well known elementary concepts on optimality conditions for convex constrained differentiable problems (for more details see e.g., \cite{B99}).
Consider a problem of the form
\begin{equation}  \mbox{(C):} \quad \min \{ g(\bx): \bx \in C\}, \end{equation}
\noindent where $C$ is a closed convex set and $g$ is a continuously differentiable function, which is possibly nonconvex.
A vector $\bx^* \in C$ is called stationary if
 \begin{equation} \label{stationary:1} \langle \nabla g(\bx^*) , \bx-\bx^* \rangle \geq 0 \mbox{ for all } \bx \in C.\end{equation}
If $\bx^*$ is an optimal solution of (P), then it is also stationary. Therefore, stationarity is a necessary condition for optimality.
Many optimization methods devised  for solving nonconvex problems of the form (C) are only guaranteed to converge to stationary points (occasionally it is only shown that all limit points of the generated sequence are stationary).

It is often useful to use the property that \textit{for any} $L>0$, a vector $\bx^*$ is a stationary point if and only if
\begin{equation} \label{xpc}  \bx^*  = P_C\left (\bx^*-\frac{1}{L} \nabla g(\bx^*) \right ),\end{equation}
where for a closed subset $D \subseteq \real^n$, the operator $P_D(\cdot)$ denotes the orthogonal projection onto $D$, that is,
$$ P_{D}(\by) \equiv \argmin_{ \bx \in D} \|\bx-\by\|^2.$$
It is interesting to note that condition (\ref{xpc}) -- although expressed in terms of the parameter $L$ --
does not actually depend on $L$ by its equivalence to (\ref{stationary:1}).\\

It is natural to try and extend  (\ref{stationary:1}) or (\ref{xpc}) to the nonconvex (feasible set) setting. Condition (\ref{stationary:1}) with $g=f$ and $C=C_s$ is actually not a necessary optimality condition so we do not pursue it further. To extend (\ref{xpc}) to the sparsity constrained problem (P), we introduce the notion of ``$L$-stationarity".
\begin{defin} A vector $\bx^*\in C_{s}$ is called an $L$-stationary point of (P) if it satisfies the relation
\begin{equation} \label{stat1} \mbox{{\rm [NC${}_L$]}} \quad \bx^* \in P_{C_s}\left ( \bx^* - \frac{1}{L} \nabla f(\bx^*) \right ).\end{equation}
\end{defin}
Note that since $C_s$ is not a convex set, the orthogonal projection operator $P_{C_s}(\cdot)$ is not single-valued.
Specifically, the orthogonal projection $P_{C_s}(\bx)$ is a vector consisting of the $s$ components of $\bx$ with the largest absolute value. In general, there could be more than one choice to the $s$ largest components. For example:
$$ P_{C_2} ((2,1,1)^T)= \left \{ (2,1,0)^T, (2,0,1)^T \right \}.$$

Below we will show that under an appropriate Lipschitz condition, $L$-stationarity is a necessary condition for optimality.  Before proving this result, we describe a more explicit representation of [NC${}_L$].
  \begin{lem} \label{lem:nc:equiv} For any $L>0$, $\bx^*$ satisfies {\rm [NC${}_L$]} if and only if $\|\bx^*\|_0 \leq s$ and
\begin{equation} \label{nc2} |\nabla_i f(\bx^*) | \left \{ \begin{array}{ll}  \leq  L M_s(\bx^*)  & \mbox{ if } i \in I_0(\bx^*),\\ =0 & \mbox{ if } i \in I_1(\bx^*).\end{array} \right.
\end{equation}
\end{lem}
\noindent {\bf Proof:} ([NC${}_L$] $\Rightarrow$ (\ref{nc2})). Suppose that $\bx^*$ satisfies [NC${}_L$]. If $i \in I_1(\bx^*)$, then by [NC${}_L$] we have $x_i^*=x_i^*-\frac{1}{L} \nabla_i f(\bx^*)$, so that $\nabla_i f(\bx^*)=0$.
If $i \in I_0(\bx^*)$, then $\left |x_i^*-\frac{1}{L} \nabla_i f(\bx^*)\right | \leq M_s(\bx^*)$, which combined with the fact that $x_i^*=0$ implies that
$|\nabla_i f(\bx^*) | \leq L M_s(\bx^*)$, and consequently $(\ref{nc2})$ holds true.\\
((\ref{nc2}) $\Rightarrow$ [NC${}_L$]). Suppose that $\bx^*$ satisfies (\ref{nc2}). If $\|\bx^*\|_0 <s$, then $M_s(\bx^*)=0$ and by (\ref{nc2}) it follows that
$\nabla f(\bx^*)=0$; therefore, in this case, $P_{C_s}\left (\bx^*-\frac{1}{L} \nabla f(\bx^*)\right )=P_{C_s}(\bx^*)$ is the set $\{\bx^*\}$.
 If $\|\bx^*\|_0=s$, then $M_s(\bx^*) \neq 0$ and $|I_1(\bx^*)|=s$. By
(\ref{nc2})
$$ \left |x_i^* - 1/L \nabla_i f(\bx^*) \right | \left \{ \begin{array}{ll} =|x_i^*|  & i \in I_1(\bx^*)\\ \leq M_s(\bx^*) & i \in I_0(\bx^*). \end{array} \right. $$
Therefore, the vector $\bx^*-\frac{1}{L} \nabla f(\bx^*)$ contains the $s$ components of $\bx^*$  with the largest absolute value and all other components are smaller or equal to them, so that
 [NC${}_L$] holds. \qed

A direct result of Lemma~\ref{lem:nc:equiv} is that any $L$-stationary point is a BF point.
\begin{corollary} \label{cor:lbfs} Suppose that $\bx^*$ is an $L$-stationary point for some $L>0$. Then $\bx^*$ is a BF point.
\end{corollary}

\begin{remark} By Lemma \ref{lem:nc:equiv} it follows that the condition for $L$-stationarity depends on $L$. In particular, [NC${}_L$] is stronger/more restrictive as
$L$ gets smaller. That is, if $\bx^*$ is an $L_1$ stationary point, then it is also an $L_2$-stationary point  for any $L_2 \geq L_1$.
 This is a different situation than the one described for problems with convex feasible sets where stationarity does not depend on any parameter. Based on this observation, it is
natural to define the \textit{stationarity level} of a BF vector  $\bx^* \in C_s$ as the smallest nonnegative $L$
 for which condition (\ref{nc2}) holds. If a BF vector $\bx^*$ satisfies $\|\bx^*\|_0<s$, then the stationarity level is zero.
 If $\|\bx^*\|_0=s$, then the stationarity level, denoted by $SL(\bx^*)$, is given by
$$ SL(\bx^*) \equiv \max_{i \in I_0(\bx^*)} \frac{|\nabla_i f(\bx^*)|}{M_s(\bx^*)}.$$
The role of SL will become apparent when we discuss the proposed algorithms.
\end{remark}

In general, $L$-stationarity is not a necessary optimality condition for problem (P). To establish such a result, we need to assume a Lipschitz continuity property of $\nabla f$.
\begin{assumption} \label{as:gl} The gradient of the objective function $\nabla f$ is Lipschitz with constant $L(f)$ over $\real^n$:
$$ \|\nabla f(\bx)-\nabla f(\by) \| \leq L(f) \|\bx-\by\| \quad \mbox{ for every } \bx,\by \in \real^n.$$
\end{assumption}
This assumption holds for $f=f_{\rm LI}$ with $L(f)=2 \lambda_{\max}(\bba^T \bba)$, but not for $f=f_{\rm QU}$. Assumption \ref{as:gl}
will \textit{not} be made throughout the paper and it will be stated explicitly when needed.

It is well known that a function satisfying Assumption \ref{as:gl} can be upper bounded
by a quadratic function whose associated matrix is a multiple of the identity matrix. This result is known as \textit{the descent lemma}:
\begin{lem}[{The Descent Lemma \cite{B99}}] \label{lem:dl} Let $f$ be  a continuously differentiable function satisfying Assumption \ref{as:gl}. Then for every $L \geq L(f)$
$$ f(\bx) \leq h_L(\bx,\by) \mbox{ for any } \bx,\by \in \real^n, $$
\noindent where
\begin{equation} \label{defh}  h_L(\bx,\by) \equiv  f(\by) + \langle \nabla f(\by), \bx-\by \rangle +\frac{L}{2} \|\bx-\by\|^2, \quad \bx,\by \in \real^n.\end{equation}
\end{lem}
Based on the descent lemma, we can prove the following technical and useful result.
\begin{lem} \label{lem:basic} Suppose that Assumption \ref{as:gl} holds and that $L > L(f)$. Then for any $\bx \in C_s$ and $\by\in \real^n$ satisfying
\begin{equation} \label{yxp} \by \in P_{C_s}\left ( \bx-\frac{1}{L} \nabla f(\bx) \right ),\end{equation}
we have
\begin{equation} \label{basic}  f(\bx)-f(\by) \geq \frac{L-L(f)}{2} \|\bx-\by\|^2.\end{equation}
\end{lem}
\noindent {\bf Proof.}  Note that (\ref{yxp}) can be written as
$$ \by \in \argmin_{\bz \in C_s} \left \|\bz-\left (\bx - \frac{1}{L}\nabla f(\bx)\right ) \right \|^2.$$
\noindent After rearrangement of terms, this minimization problem can be easily seen to be equivalent to
$$ \by \in \argmin_{\bz \in C_s} h_L(\bz,\bx).$$
This implies that
\begin{equation} \label{hxyp} h_L(\by,\bx) \leq h_L(\bx,\bx) =f(\bx).\end{equation}
Now, by the descent lemma we have
$$ f(\bx)-f(\by) \geq f(\bx)-h_{L(f)}(\by,\bx), $$
\noindent which combined with (\ref{hxyp}) and the identity  $$ h_{L(f)}(\bx,\by) =h_L(\bx,\by)-\frac{L-L(f)}{2}\|\bx-\by\|^2,$$
\noindent yields (\ref{basic}). \qed

Under Assumption \ref{as:gl} we now show that an optimal solution of (P) is an $L$-stationary point for any $L > L(f)$.
\begin{theorem} \label{thm:lstat} Suppose that Assumption \ref{as:gl} holds, $L > L(f)$ and let $\bx^*$ be an optimal solution of (P). Then
\begin{itemize}
\item[(i)] $\bx^*$ is an $L$-stationary point.
\item[(ii)] The set $P_{C_s}\left ( \bx^* - \frac{1}{L} \nabla f(\bx^*) \right )$ is a singleton\footnote{A set is called a \textit{singleton} if it contains exactly one element.}.
\end{itemize}
\end{theorem}
\noindent {\bf Proof:} We will prove both parts simultaneously. Suppose to the contrary that there exists a vector
\begin{equation}
\by \in P_{C_s} \left ( \bx^*-\frac{1}{L} \nabla f(\bx^*) \right ),
\end{equation}
which is different from $\bx^*$ ($\by \neq \bx^*$). Invoking Lemma \ref{lem:basic} with $\bx=\bx^*$, we have
$$ f(\bx^*)-f(\by) \geq \frac{L-L(f)}{2} \|\bx^*-\by\|^2,$$
\noindent contradicting the optimality of $\bx^*$.  We conclude that $\bx^*$ is the only vector \sloppy in the set  $P_{C_s} \left ( \bx^*-\frac{1}{L} \nabla f(\bx^*) \right )$. \qed

To conclude this section, we have shown that under a Lipschitz condition on $\nabla f$, $L$-stationarity for any $L>L(f)$ is a necessary optimality condition, which also implies the basic feasibility property. In Section~\ref{sec:ssparse} we will show how the iterative hard thresholding method for solving the general problem (P), can be used in order to find $L$-stationary points (for $L>L(f)$).

\subsection{Coordinate-Wise Minima}

 The $L$-stationarity necessary optimality condition has two major drawbacks: first, it requires the function's gradient to be Lipschitz continuous and second, in order to validate it, we need to know a bound on the Lipschitz constant. We now consider a different and stronger necessary optimality condition that does not require such knowledge on the Lipschitz constant, and in fact does not even require Assumption \ref{as:gl} to hold.

For a general unconstrained optimization problem, a vector $\bx^*$ is called a ``coordinate-wise (CW)" minimum if for every $i=1,2,\ldots,n$ the scalar $x_i^*$ is a minimum
 of $f$
with respect to the $i$th component $x_i$ while keeping all other variables fixed:
$$ x_i^* \in \argmin f(x_1^*,\ldots,x_{i-1}^*, x_i,x_{i+1}^*, \ldots,x_n^*).$$
Clearly, any optimal $\bx^*$ is also a coordinate-wise minimum.
It is therefore natural to extend this definition to problem (P), in order to obtain an alternative necessary condition.
\begin{defin} Let $\bx^*$ be a feasible solution of (P). Then $\bx^*$ is called  a coordinate-wise (CW) minimum of (P) if one of the following cases hold true:\\
Case I: $\|\bx^*\|_0 <s$ and for every $i=1,2,\ldots,n$ one has:
\begin{equation} \label{cw-case1} f(\bx^*) = \min_{t \in \real} f(\bx^*+t\be_i).\end{equation}
Case II: $\|\bx^*\|_0=s$ and for every $i \in I_1(\bx^*)$ and $j=1,2,\ldots,n$ one has:
\begin{equation} \label{cw-case2}  f(\bx^*) \leq \min_{t \in \real} f(\bx^*-x_i^*\be_i+t \be_j).\end{equation}
\end{defin}

Obviously, any optimal solution of (P) is also a CW-minimum. This is formally stated in the next theorem.
\begin{theorem} \label{the:optcw} Let $\bx^*$ be an optimal solution of (P). Then $\bx^*$ is a CW-minimum of (P). \end{theorem}

It is easy to see that any CW-minimum is also a BF vector, as stated in the following lemma.
\begin{lem} \label{lem:cw-bfs} Let $\bx^* \in C_s$ be a CW-minimum of (P). Then $\bx^*$ is also a BF vector.
\end{lem}
\noindent {\bf Proof.}
We first show that if a vector $\bx^*$ satisfying $\|\bx^*\|_0=s$ is a CW-minimum of (P), then (\ref{cw-case1}) is satisfied for any
$i \in I_1(\bx^*)$. Indeed, let $i \in I_1(\bx^*)$ and take $j=i$. Then (\ref{cw-case2}) becomes
\begin{equation}
\label{eq:cwp}
f(\bx^*) \leq \min_{t \in \real} f(\bx^*-x_i^*\be_i+t \be_i).
\end{equation}
Since $f(\bx^*-x_i^*\be_i+x_i^*\be_i)=f(\bx^*)$, it follows that (\ref{eq:cwp}) is equivalent to
$$  f(\bx^*) = \min_{t \in \real} f(\bx^*-x_i^*\be_i+t \be_i),$$
\noindent which letting $s=t-x_i^*$ becomes
$$  f(\bx^*) = \min_{s \in \real} f(\bx^*+s \be_i).$$

We conclude that for any CW-minimum $\bx^*$ of (P) we have
\begin{equation} \label{reim} \nabla_i f(\bx^*) =0 \mbox{ for all } i \in I_1(\bx^*).\end{equation}
In addition, in case I we obviously have that $\nabla f(\bx^*)=0$, which completes the proof. \qed

We have previously established under Assumption \ref{as:gl} in Theorem \ref{thm:lstat} that being an $L$-stationary point for $L>L(f)$ is a necessary condition for optimality.
 A natural question that arises is what is the relation between CW-minima and $L$-stationary points (for $L>L(f)$). We will show that being a CW-minimum is a stronger, i.e. more restrictive,
  condition than being an $L$-stationary point for any $L \geq L(f)$. In fact, a stronger result will be established: any CW-minimum
  is also an $\tilde{L}$ stationary point for an $\tilde{L}$ which is less than or equal to $L(f)$. In practice, $\tilde{L}$ can be much smaller than $L(f)$.

  In order to precisely define $\tilde{L}$, we note that under Assumption \ref{as:gl}, it follows immediately that for any $i \neq j$ there exists a constant $L_{i,j}(f)$ for which
\begin{equation} \label{local-lipschitz}  \|\nabla_{i,j}f(\bx) - \nabla_{i,j}f(\bx+\bd)\| \leq L_{i,j}(f) \|\bd\|,\end{equation}
\noindent for any $\bx \in \real^n$ and any $\bd \in \real^n$ which has at most two nonzero components. Here $\nabla_{i,j}f(\bx)$ denotes a vector of length-$2$ whose elements are the $i$th and $j$th elements of $\nabla f(\bx)$.  We will be especially interested in the following constant, which we call \textit{the local Lipschitz constant}:
$$ L_2(f) \equiv \max_{i \neq j} L_{i,j}(f).$$
Clearly (\ref{local-lipschitz}) is satisfied when replacing $L_{i,j}(f)$ by $L(f)$. Therefore, in general,
$$ L_2(f) \leq L(f).$$
In practice, $L_2(f)$ can be much smaller than $L(f)$ as the following example illustrates.
\begin{example} \label{example:locallips} Suppose that the objective function in (P) is $f(\bx)=\bx^T \bbq \bx+2 \bb^T \bx$, with $\bb$ being a  vector in $\real^n$ and
$$ \bbq  = \bbi_n + \bbj_n,$$
\noindent where $\bbi_n$ is the $n \times n$ identity matrix and $\bbj_n$ is the $n \times n$ matrix of all ones. Then
$$ L(f) = 2 \lambda_{\max}(\bbq) = 2 \lambda_{\max}(\bbi_n+\bbj_n) = 2 (n+1).$$
On the other hand, for any $i \neq j$ the constant $L_{i,j}(f)$ is twice the maximum eigenvalue of the submatrix of $\bbq$ consisting of the $i$th and $j$th rows and columns. That is,
$$ L_{i,j}(f) = 2 \lambda_{\max}\begin{pmatrix} 2 & 1 \\ 1 & 2 \end{pmatrix} = 6.$$
For large $n$, $ L(f) = 2n+2$  can be much larger than $L_2(f)=6.$
\end{example}
It is not difficult to see that the descent lemma (Lemma~\ref{lem:dl}) can be refined to a suitable ``local" version.
\begin{lem}[Local Descent Lemma] \label{lem:localdescent} Suppose that Assumption \ref{as:gl} holds. Then
$$ f(\bx+\bd) \leq f(\bx)+\nabla f(\bx)^T \bd +\frac{L_2(f)}{2}\|\bd\|^2$$
for any vector $\bd \in \real^n$ with at most two nonzero components.
\end{lem}
Using the local descent lemma we can now show that a CW-minimum is also an $L_2(f)$-stationary point.
\begin{theorem} \label{the:cwlw} Suppose that Assumption \ref{as:gl} holds and let $\bx^*$ be a CW-minimum of (P). Then
\begin{equation} \label{nc-cw} |\nabla_i f(\bx^*)| \left \{ \begin{array}{ll} \leq L_2(f) M_s(\bx^*) & i \in I_0(\bx^*), \\  =0 & i \in I_1(\bx^*), \end{array} \right.\end{equation}
that is, $\bx^*$ is an $L_2(f)$-stationary point.
\end{theorem}
\noindent {\bf Proof:} Since $\bx^*$ is a CW-minimum, it follows by Lemma \ref{lem:cw-bfs} that it is a BF vector. Thus, since $\|\bx^*\|_0<s$, we have $\nabla f(\bx^*)=\bo$, establishing the result for this case. \\

Suppose now that $\|\bx^*\|_0=s$. Let $i \in I_1(\bx^*)$. Then, again,  by Lemma \ref{lem:cw-bfs} it follows that $\bx^*$ is a BF vector and thus $\nabla_i f(\bx^*)=0$.
Now let $i \in I_0(\bx^*)$ and let $m$ be an index for which $|x_m^*|=M_s(\bx^*)$. Obviously, $m \in I_1(\bx^*)$, and thus, since $\bx^*$ is a CW-minimum, it follows in particular that
\begin{equation} \label{a1}  f(\bx^*) \leq f(\bx^*-x_m^*\be_m-\sigma x_m^* \be_i),\end{equation}
\noindent where $\sigma = \sgn(x_m^* \nabla_i f(\bx^*))$.  By the local descent lemma (Lemma \ref{lem:localdescent})  we have
\begin{eqnarray} \nonumber  f(\bx^*-x_m^*\be_m-\sigma x_m^* \be_i) &\leq& f(\bx^*) + \nabla f(\bx^*)^T (-x_m^* \be_m-\sigma x_m^*\be_i)+\frac{L_2(f)}{2} \|x_m^* \be_m+\sigma x_m^* \be_i\|^2\\
\nonumber &=& f(\bx^*) -x_m^* \nabla_m f(\bx^*) -\sigma x_m^* \nabla_i f(\bx^*) + L_2(f) (x_m^*)^2\\
&=& f(\bx^*) -\sigma x_m^* \nabla_i f(\bx^*) + L_2(f) (x_m^*)^2, \label{a2}
\end{eqnarray}
\noindent where the last equality follows from the fact that since $m \in I_1(\bx^*)$, it follows by (\ref{reim}) that  $\nabla_m f(\bx^*)=0$.\\
\indent Combining (\ref{a1}) and (\ref{a2}) we obtain that
$$ 0 \leq -\sigma x_m^* \nabla_i f(\bx^*) +L_2(f) (x_m^*)^2.$$
Recalling the definition of $\sigma$, we conclude that
$$ |x_m^* \nabla_i f(\bx^*)| \leq L_2(f) (x_m^*)^2,$$
\noindent which is equivalent to
$$ | \nabla_i f(\bx^*)| \leq L_2(f) |x_m^*| = L_2(f)M_s(\bx^*),$$
concluding the proof.
\qed

An immediate consequence of Theorem~\ref{the:cwlw} is that under Assumption \ref{as:gl}, any optimal solution of (P) is an $L_2(f)-$stationary point.
\begin{corollary} \label{cor:optl2} Suppose that Assumption \ref{as:gl} holds. Then any optimal solution of (P) is also an $L_2(f)-$stationary point of (P).
\end{corollary}

To summarize our discussion on optimality conditions we have shown that without Assumption \ref{as:gl} we have the following relations:
$$\begin{array}{rc} & \mbox{ optimal solution of (P)} \\  \mbox{ Theorem ~\ref{the:optcw}} & \Downarrow\\  & \mbox{ CW-minimum of (P) } \\ \mbox{Lemma~\ref{lem:cw-bfs}} & \Downarrow \\ & \mbox{ BF vector of (P)}\end{array}$$
Under Assumption \ref{as:gl}, we have:
$$\begin{array}{lc} & \mbox{ optimal solution of (P)} \\ \mbox{ Theorem~\ref{the:optcw}} & \Downarrow \\ & \mbox{ CW-minimum of (P) } \\ \mbox{Theorem~\ref{the:cwlw}} & \Downarrow \\ & L_2(f)-\mbox{stationary} \\ \mbox{Corrolary~\ref{cor:lbfs}} & \Downarrow \\
&  \mbox{ BF vector  of (P)}\end{array}$$

To illustrate these relationships we consider a detailed example.
\begin{example} Consider problem (P) with $s=2, n=5$ and
$$ f(\bx)= \bx^T \bbq \bx +2 \bb^T \bx,$$
\noindent where $\bbq = \bbi_5+\bbj_5$ as in Example \ref{example:locallips}, and $\bb= -(3,2,3,12,5)^T$. In Lemma~\ref{lem:lfinite} we showed how to compute the BF vectors of problem (P) with a quadratic objective. Using this method it is easy to see that in our case there are 10 BF vectors
given by (each corresponding to a different choice of two variables out of 5):
\begin{eqnarray*} \bx_1 &=& ( 1.3333  ,  0.3333,         0   ,      0   ,      0)^T,\\
\bx_2 &=&(   1.0000     ,    0   , 1.0000      ,   0     ,    0)^T,\\
\bx_3 &=&(  -2.0000  ,       0    ,     0   , 7.0000        , 0)^T,\\
\bx_4 &=&(   0.3333  ,       0     ,    0    ,     0  ,  2.3333)^T,\\
\bx_5 &=&(        0  ,  0.3333  ,  1.3333   ,      0     ,    0)^T,\\
\bx_6 &=&(        0  , -2.6667     ,    0   , 7.3333    ,     0)^T,\\
\bx_7 &=&(        0  , -0.3333    ,     0   ,      0   , 2.6667)^T,\\
\bx_8    &=&(     0   ,      0   ,-2.0000   , 7.0000    ,     0)^T,\\
\bx_9        &=&( 0    ,     0   , 0.3333   ,      0   , 2.3333)^T,\\
\bx_{10}    &=&(     0    ,     0  ,       0   , 6.3333 ,  -0.6667)^T.
 \end{eqnarray*}

 The stationarity levels and function values up to two digits of accuracy of each of the BF vectors  is given in Table~\ref{table:funstat1}.
\begin{table}[h]
\begin{center}
\begin{tabular}{||c||c|c|c|c|c|c|c|c|c|c||}\hline\hline
BF vector number & 1 & 2 & 3 & 4 & 5 & 6 & 7 & 8 & 9 & 10\\\hline
function value  & -4.66 &  -6.00&  -78 & -12.66 &  -4.66 & -82.66&  -12.66 & -78 & -12.66 & -72.66\\
stationarity level &  62  & 20   & 3  & 56   &62 &   1.25  & 58  &  3   &56&   11\\
\hline \hline
\end{tabular}
\caption{Function values and stationarity levels of the 10 BF vectors.}\label{table:funstat1}
\end{center}
\end{table}

Since in this case $L_2(f)=6$ (see Example \ref{example:locallips}), it follows by Corollary \ref{cor:optl2} that any optimal solution is a $6$-stationary point, implying that only the three BF vectors  $\bx_3,\bx_6,\bx_8$ are candidates for being optimal solutions. In addition, by Theorem \ref{the:cwlw}, only these three BF vectors may be CW-minima. By direct calculation we found that only $\bx_6$ -- the optimal solution of the problem -- is a CW-minima. Therefore, in this case, the only CW-minima is the global optimal solution. Note, however, that there could of course be examples in which there exist CW-minima which are not optimal.
\end{example}

\section{Numerical Algorithms}
\label{sec:alg}

We now develop two classes of algorithms that achieve the necessary conditions defined in the previous section:
 \begin{itemize} \item {\bf Iterative hard thresholding (IHT).} The first algorithm, results from using the $L$-stationary condition. For the case $f \equiv f_{\rm LI}$, and under the assumption that $\|\bba\|_2<1$, it coincides with the IHT method \cite{BD08}. Our approach extends this algorithm to the general case under Assumption \ref{as:gl}, and it will be refereed to as ``the IHT method" in our general setting as well. We will prove that the limit points of the algorithm are $L(f)-$stationary points. As we show, this method is well-defined only when Assumption \ref{as:gl} holds and relies on knowledge of the Lipschitz constant.
 \item {\bf Sparse-simplex methods.} The other two algorithms we suggest are essentially coordinate descent methods that optimize the objective function at each iteration with respect to either one or two decision variables.

      The first algorithm in this class seeks the coordinate, or coordinates, that lead to the largest decrease and optimizes with respect to them. Since the support of the iterates changes by at most \textit{one} index, it has some resemblance to the celebrated simplex method for linear programming and will thus be referred to as ``the greedy sparse-simplex method". We show that any limit point of the sequence generated by this approach is a CW-minima, which as shown in Theorem \ref{the:cwlw}, is a stronger notion than $L-$stationarity for any $L\geq L_2(f)$. An additional advantage of this approach is that it is well defined even when Assumption \ref{as:gl} is not valid, and does not require any knowledge of the Lipschitz constant even  when one exists. The disadvantage of the greedy sparse-simplex method is that it does not have a selection strategy for choosing the indices of the variables to be optimized, but rather explores all possible choices. Depending on the objective, this may be a very costly step.

     To overcome this drawback, we suggest a second coordinate descent algorithm with an extremely simple index selection rule; this rule discards the need to perform an exhaustive search for the relevant indices on which the optimization will be performed. This approach will be refereed to as ``the partial sparse-simplex method". Under Assumption \ref{as:gl} we show that it is guaranteed to converge to $L_2(f)$-stationary points.
 \end{itemize}

In the ensuing subsections we consider each of the algorithms above. We present the methods along with statements regarding their convergence properties. The detailed proofs of the convergence results are deferred to Section~\ref{sec:proof}.

\subsection{The IHT Method}
\label{sec:ssparse}
\indent One approach for solving problem (P) is to employ the following fixed point method in order to ``enforce" the $L$-stationary condition
(\ref{stat1}):
\begin{equation} \label{gp} \bx^{k+1} \in P_{C_s}\left (\bx^k-\frac{1}{L}    \nabla f(\bx^k) \right ), \quad k=0,1,2,\ldots\end{equation}
Convergence results on this method can be obtained when Assumption \ref{as:gl} holds; we will therefore make this assumption throughout this subsection. The iterations defined by (\ref{gp}) were studied in \cite{BD08} for the special case in which $f\equiv f_{\rm LI}$ and $\|\bba\|_2<1$, and were referred to as the ``$M$-sparse" algorithm. Later on, in \cite{BD10}, the authors referred to this approach as the IHT method (again, for $f=f_{\rm LI}$) and analyzed a version with an adaptive step size which avoids the need for the normalization property $\|\bba\|_2<1$.  Similarly, we refer to this approach for more general obejctvie functions as the IHT method:
\fcolorbox{black}{Ivory2}{\parbox{15cm}{{\bf The IHT method}\\
\medskip
{\bf Input:} a constant $L\geq L(f)$. \\
$\bullet$ {\bf Initialization:} Choose $\bx_0 \in C_s$.\\
$\bullet$ {\bf General step :} $  \bx^{k+1} \in P_{C_s} \left ( \bx^k - \frac{1}{L} \nabla f(\bx^k) \right ), \quad (k=0,1,2,\ldots)$
}}
\smallskip

It can be shown that the general step of the IHT method is equivalent to the relation
\begin{equation} \label{gph} \bx^{k+1} \in \argmin_{\bx \in C_s} h_L(\bx,\bx^k),\end{equation}
where $h_L(\bx,\by)$ is defined by (\ref{defh}) (see also the proof of Theorem \ref{thm:lstat}).

%
%

Several basic properties of the IHT method are summarized in the following lemma.
\begin{lem} \label{lem:sdescent} Let $\{\bx^k\}_{k \geq 0}$ be the sequence generated by the IHT  method with a constant stepsize $\frac{1}{L}$ where $L > L(f)$. Then
\begin{enumerate}
\item $f(\bx^k)-f(\bx^{k+1}) \geq \frac{L-L(f)}{2} \|\bx^k-\bx^{k+1}\|^2.$
\item $\{f(\bx^k)\}_{k \geq 0}$ is a nonincreasing sequence.
\item $\|\bx^k-\bx^{k+1}\| \rightarrow 0$.
\item For every $k=0,1,2,\ldots,$ if $\bx^k \neq \bx^{k+1}$, then $f(\bx^{k+1})<f(\bx^k).$
\end{enumerate}
\end{lem}
\noindent {\bf Proof:} Part 1 follows by substituting $\bx=\bx^{k},\by = \bx^{k+1}$ in (\ref{basic}). Parts 2,3 and 4 follow immediately from part 1. \qed

A direct consequence of Lemma \ref{lem:sdescent} is the convergence of the sequence of function values.

\begin{corollary} Let $\{\bx^k\}_{k \geq 0}$ be the sequence generated by the IHT method with a constant stepsize $\frac{1}{L}$ where $L > L(f)$. Then the sequence $\{f(\bx^k)\}_{k \geq 0}$ converges.
\end{corollary}

As we have seen, the  IHT algorithm can be viewed as a fixed point method for solving the condition for $L$-stationarity. The following theorem states that all accumulation points of the sequence generated by the IHT  method with constant stepsize $\frac{1}{L}$ are
indeed $L$-stationary points.
\begin{theorem} \label{thm:cong} Let $\{\bx^k\}_{k \geq 0}$ be the sequence generated by the IHT method with stepsize $\frac{1}{L}$ where $L>L(f)$. Then any accumulation point of $\{\bx^k\}_{k \geq 0}$ is an $L$-stationary point.
\end{theorem}
\noindent {\bf Proof.} See Section \ref{sec:thm:cong}.

\subsubsection{The Case $f=f_{\rm LI}$}
When $f(\bx)\equiv f_{\rm LI}(\bx) \equiv \| \bba \bx-\bb\|^2,$ and under the assumption of $s$-regularity of (P), we know by Lemma \ref{lem:lfinite} that the number of BF vectors is finite. Utilizing this fact we can now show convergence of the whole sequence generated by the IHT method when $f = f_{\rm LI}$. This result is stronger than the one of Theorem \ref{thm:cong}, which only shows that all accumulation points are $L$-stationary points.
\begin{theorem} \label{the:conver} Let $f(\bx) \equiv f_{\rm LI}(\bx)=\|\bba\bx-\bb\|^2$. Suppose that the $s$-regularity property holds for the matrix $\bba$. Then the sequence generated by the
IHT method with stepsize $\frac{1}{L}$  where $L>L(f)$ converges to an $L$-stationary point.
\end{theorem}
\noindent {\bf Proof.} See Section \ref{sec:the:conver}.

\begin{remark}
As we have noted previously, the IHT method in the case $f = f_{\rm LI}$ with fixed step-size set to $1$ was proposed in \cite{BD08}.  It was shown in  \cite{BD08} that if $\bba$ satisfies the $s$-regularity property and  $\|\bba\|_2 < 1$, then the
algorithm converges to a local minimum. This result is consistent with Theorem~\ref{the:conver} since when $\|\bba\|_2<1$, the Lipschitz
constant satisfies $L(f)<1$, and we can therefore assure convergence by Theorem \ref{the:conver} with stepsize equal to 1.
In \cite{BD10} the authors note that the IHT method with stepsize 1 might diverge when $\|\bba\|_2>1$. To overcome this limitation, they propose an adaptive stepsize for which they show the same type of convergence results. Our result here shows that a fixed step size which depends on the Lipschitz constant can also be used.
\end{remark}

\subsubsection{Examples}
\begin{figure}[]
$$\begin{array}{cc}
\mbox{L=250} & \mbox{L=500}\\
\includegraphics[scale=0.5]{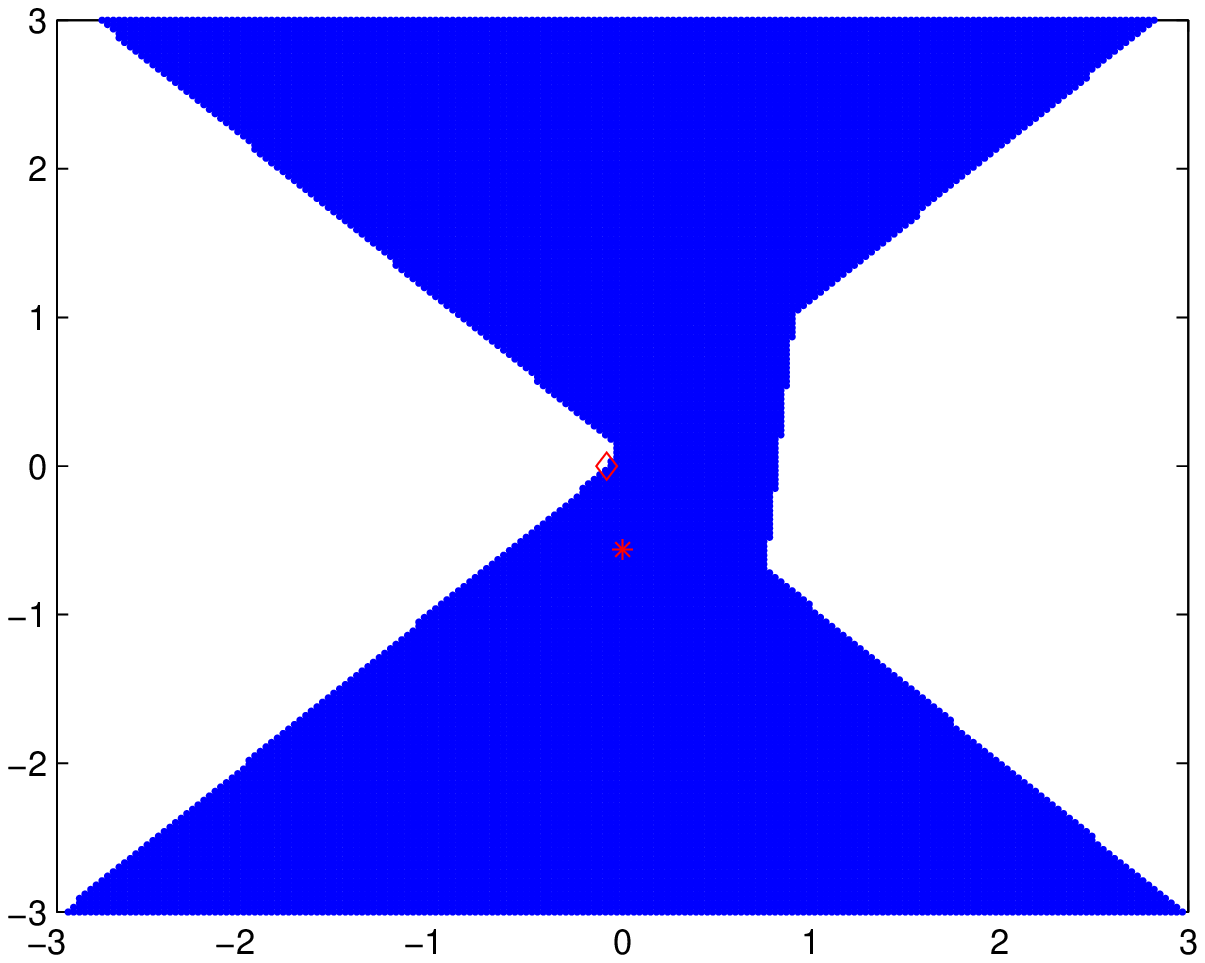} & \includegraphics[scale=0.5]{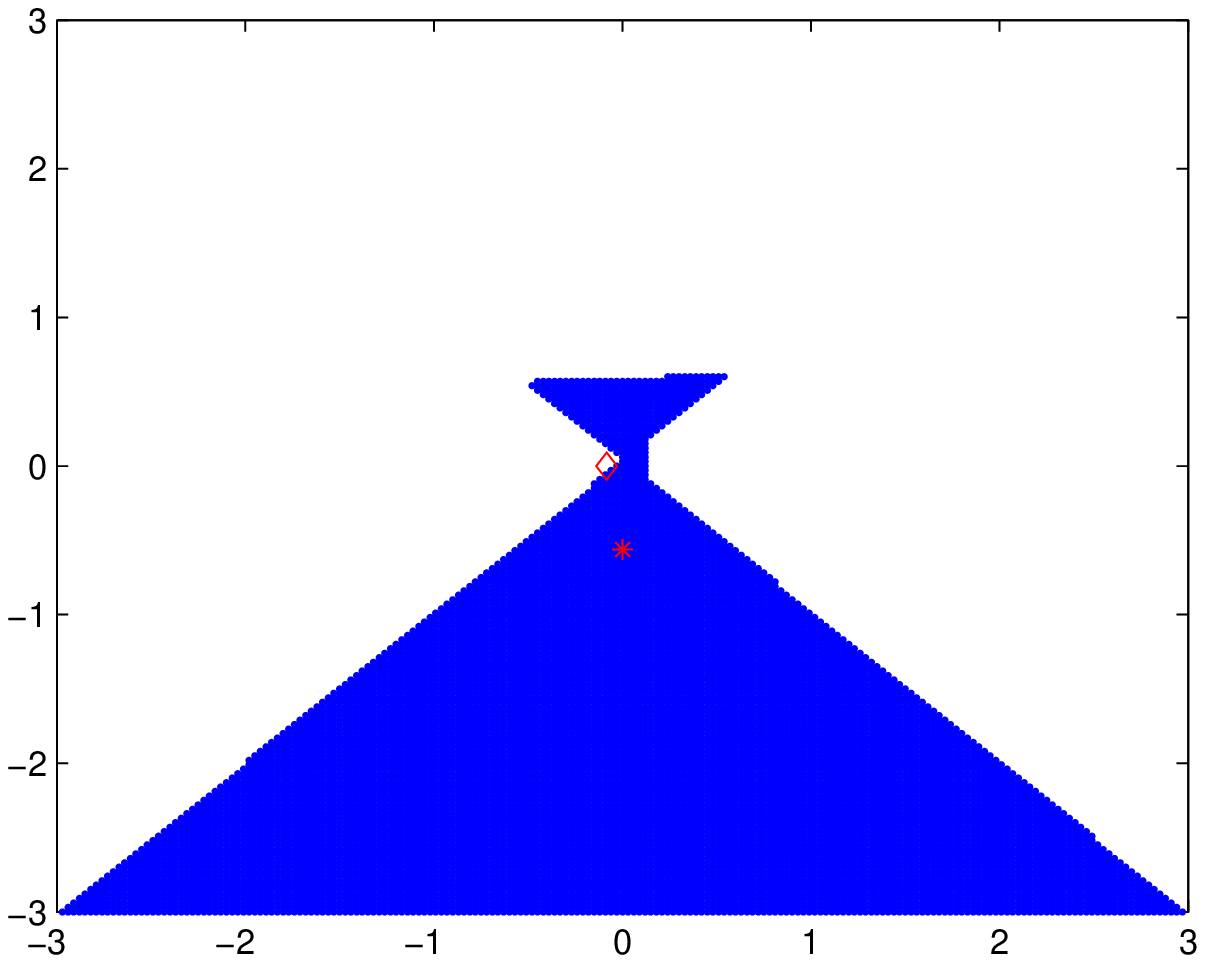}
\end{array} $$
\label{fig:L250500}
\caption{The optimal solution $(0,-9/16)^T$ is denoted by a red asterisk and the additional BF vector $(-1/12,0)^T$ is denoted by a red diamond. The region of convergence to the optimal solution is the blue region and the points in the white region converged to the non-optimal point $(-1/12,0)^T$. The left image describes the convergence region when the IHT method was invoked with $L=250$, while the right image describes the same for $L=500$. When $L$ gets larger, the chances to converge to the non-optimal $L$-stationary point are higher.}
\end{figure}

\begin{example} Consider the problem
\begin{equation} \label{ex1}  \min \left \{ f(x_1,x_2) \equiv 12x_1^2+20x_1x_2+32x_2^2 : \left \|(x_1;x_2)^T \right \|_0 \leq 1 \right \}.\end{equation}
The objective function is convex quadratic and the Lipschitz constant of its gradient is given by
$$ L(f)= 2 \lambda_{\max}\begin{pmatrix} 12 & 10 \\ 10 & 16 \end{pmatrix} = 48.3961.$$
It can be easily seen that there are only two BF vectors to this problem: $(0,-9/16)^T,(-1/12,0)^T$ (constructed by taking one variable to be zero and the other to satisfy that the corresponding partial derivative is zero). The optimal solution of the problem is the first BF vector $(0,-9/16)^T$ with objective function value of $-81/16$. This point is an $L$-stationary point for any $L \geq L(f)$. The second point $(-1/12,0)^T$ is not an optimal solution (its objective function value is $-1/12$). Since $\nabla_2 f((-1/12,0)^T) = 49/3$, it follows by Lemma \ref{lem:nc:equiv} that it is an $L$-stationary point for $L\geq \frac{49/3}{1/12}=196$. Therefore, for any $L\in [L(f),196)$, only the optimal solution $(0,-9/16)^T$ is an $L$-stationary point and the IHT method is guaranteed to converge to the global optimal solution. However, if the upper bound is chosen to satisfy $L \geq 196$,  then $(-1/12,0)^T$ is also an $L$-stationary point and the IHT method might converge to it. This is illustrated in Figure \ref{fig:L250500}.
\end{example}

\begin{example} For any two positive number $a<b$, consider the problem
$$ \min \{ f(x_1,x_2) \equiv a(x_1-1)^2+b(x_2-1)^2 : \|(x_1,x_2)^T\|_0 \leq 1\}.$$
Obviously the optimal solution of the problem is $(x_1,x_2)=(0,1)$. An additional BF vector is  $\tilde{\bx}=(1,0)^T$. Note that here $L(f)=2b$. Therefore, since  $\nabla f(\tilde{\bx})=(0,-2b)^T$ and $M_1(\tilde{\bx}) = 1$, it follows that
$$ |\nabla_2f(\tilde{\bx})| \leq L(f)M_1(\tilde{\bx}),$$
\noindent and hence $\tilde{\bx}$ will also be an $L$-stationary point \textit{for any} $L \geq L(f)$. Therefore, in this problem, regardless of the value of $L$, there is always a chance to converge to a non-optimal solution.
\end{example}

\subsection{The Greedy Sparse-Simplex Method}

The IHT method is able to find $L$-stationary points for any $L>L(f)$ under Assumption \ref{as:gl}. However, by Corollary \ref{cor:optl2},
any optimal solution is also an $L_2(f)$-stationary point, and $L_2(f)$ can be significantly smaller than $L(f)$.
 It is therefore natural to seek a  method
that is able to generate such points. An even better approach would be to derive an algorithm that converges to a  CW-minima, which  by Theorem \ref{the:cwlw}, is  a stronger notion than $L$-stationary.
An additional drawback of the IHT method is that it requires the validity of Assumption \ref{as:gl} and the knowledge of the Lipschitz constant $L(f)$.\\

Below we present {\it the greedy sparse-simplex method} which overcomes the faults of the IHT method alluded to above:
    its limit points are  CW-minima, it does not require the validity of Assumption \ref{as:gl}, but if the assumption does hold, than its limit points are $L_2(f)-$stationary points (without the need to know any information on Lipschitz constants).

\fcolorbox{black}{Ivory2}{\parbox{15cm}{{\bf The Greedy Sparse-Simplex Method}\\

\medskip
$\bullet$ {\bf Initialization:} Choose $\bx_0 \in C_s$.\\
$\bullet$ {\bf General step : ($k=0,1,\ldots$)}
\begin{itemize} \item  If $\|\bx^k\|_0<s$, then compute for every $i=1,2,\ldots,n$
\begin{eqnarray}
\label{m1} t_i &\in& \argmin_{t \in \real} f(\bx^k+t \be_i),\\
\nonumber f_i &=& \min_{t \in \real} f(\bx^k+t \be_i).
\end{eqnarray}
Let $i_k \in \argmin_{i=1,\ldots,n} f_i$. If $f_{i_k}<f(\bx^k)$, then set $$\bx^{k+1} = \bx^k+t_{i_k} \be_{i_k}.$$
Otherwise, STOP.
\item If $\|\bx^k\|_0=s$, then for every $i \in I_1(\bx^k)$ and $j=1,\ldots,n$ compute
\begin{eqnarray}
\label{m2}t_{i,j} &\in & \argmin_{t \in \real} f(\bx^k-x_i^k \be_i +t \be_j),\\
\nonumber f_{i,j} &=& \min_{t \in \real} f(\bx^k-x_i^k\be_i+t\be_j).
\end{eqnarray}
Let $(i_k,j_k) \in \argmin \{ f_{i,j} : i \in I_1(\bx^k), j=1,\ldots,n\}.$ If $f_{i_k,j_k}<f(\bx^k)$, then set
$$ \bx^{k+1} = \bx^k-x^k_{i_k}\be_{i_k}+t_{i_k,j_k} \be_{j_k}.$$
Otherwise, STOP.
\end{itemize}
}}
\medskip

\begin{remark} \label{remark:greedycubic} One advantage of the greedy sparse-simplex method is that it can be easily implemented for the case $f\equiv f_{\rm QU}$, that is, the case when the objective function is quartic.
In this case the minimization steps (\ref{m1}) and (\ref{m2}) consist of finding the minimum of a scalar quartic (though nonconvex) function, which is an easy task since the minimizer is one of the at most three roots of
the cubic polynomial derivative.
\end{remark}

By its definition, the greedy sparse-simplex  method generates a non-increasing sequence of function values and gets stuck only at CW-minima.
\begin{lem} \label{lem:bpgls} Let $\{\bx^k\}$ be the sequence generated by the greedy sparse-simplex method. Then
$f(\bx^{k+1}) \leq f(\bx^k)$ for every $k\geq 0$ and equality holds if and only if $\bx^k=\bx^{k+1}$ and $\bx^k$ is a CW-minimum.
\end{lem}

Theorem~\ref{the:gscw} establishes the main convergence result for the greedy simplex-sparse method, namely that its accumulation points are CW-minima.
\begin{theorem} \label{the:gscw} Let $\{\bx^k\}$ be the sequence generated by the greedy sparse-simplex method. Then any accumulation point of $\{\bx^k\}$ is a CW-minimum of (P).
\end{theorem}
\noindent {\bf Proof.} See Section \ref{sec:the:gscw}.

Combining Theorem \ref{the:gscw} with Theorem \ref{the:cwlw} leads to the following corollary.

\begin{corollary} Suppose that Assumption \ref{as:gl} holds and let $\{\bx^k\}$ be the sequence generated by the greedy sparse-simplex method.
Then any accumulation point of $\{\bx^k\}$ is an $L_2(f)$-stationary point of (P).
\end{corollary}

\subsubsection{The Case $f=f_{\rm LI}$}
\label{subsec:mp}
We consider now the greedy sparse-simplex method when $f\equiv f_{\rm LI}$.
At step (\ref{m1}) we perform the minimization $t_i=\arg \min f(\bx^k+t\be_i)$. Since $f(\bx^k+t\be_i)=\|\bba\bx^k-\bb+t \ba_i\|^2$ ($\ba_i$ being the $i$th column of $\bba$),
 we have immediately that
$$t_i=-\frac{\ba_i^T \br_k}{\|\ba_i\|^2},$$
where $\br_k=\bba\bx^k-\bb$. We can then continue to compute
\begin{equation}
\label{eq:fi}
f_i=\left\|\br_k- \frac{\ba_i^T \br_k}{\|\ba_i\|^2}\ba_i\right\|^2=
\|\br_k\|^2-\frac{(\ba_i^T \br_k)^2}{\|\ba_i\|^2},
\end{equation}
so that
$$
i_k \in \argmin_{i=1,\ldots,n} f_i=\argmax_{i=1,\ldots,n}  \frac{|\ba_i^T \br_k|}{\|\ba_i\|}.
$$

The algorithm then proceeds as follows. For $\|\bx^k\|_0<s$ we choose
\begin{equation}
\label{eq:maxm}
i_k \in \argmax_{i=1,\ldots,n}  \frac{|\ba_i^T \br_k|}{\|\ba_i\|}.
\end{equation}
If $\ba_{i_k}^T\br_k \neq 0$, then we set
$$
\bx^{k+1}=\bx^k-\frac{\ba_{i_k}^T \br_k}{\|\ba_{i_k}\|^2}\be_{i_k}.
$$
In this case,
$$
\br_{k+1}=\bba\bx^{k+1}-\bb=\br_k-\frac{\ba_{i_k}^T \br_k}{\|\ba_{i_k}\|^2}\ba_{i_k}.
$$
Otherwise we stop. Note, that if $\bba$ has full row-rank, then
$\ba_{i_k}^T\br_k = 0$ only if $\br_k=0$. \\
For $\|\bx^k\|_0=s$ we choose
$$
(i_k,j_k)=\argmax_{i \in I_1(\bx^k), j \in \{1,2,\ldots,n\}}  \frac{|\ba_j^T \br_k^i|}{\|\ba_j\|},
$$
with $\br_k^i=\bba\bx^k-x_i^k \ba_i-\bb$. Let
$f_{i_k,j_k}=f(\bx^k-x_{i_k}^k\be_{i_k}+t\be_{j_k})$ with
$$
t=-\frac{\ba_{j_k}^T \br_k^{i_k}}{\|\ba_{j_k}\|^2}.
$$
If $f_{i_k,j_k}<f(\bx^k)$, then we set
$$
\bx^{k+1}=\bx^k-x_{i_k}^k\be_{i_k}-\frac{\ba_{j_k}^T \br_k^{i_k}}{\|\ba_{j_k}\|^2}\be_{j_k}.
$$
Otherwise we stop.

It is interesting to compare the resulting iterations with the matching-pursuit (MP) algorithm \cite{MZ93} designed to find a sparse solution to the system $\bba \bx = \bb$. The MP method begins with an initial guess of $\bx^0=\bf{0}$ and $\br_0=\bb$.
 At each iteration, we add an element to the support by choosing
\begin{equation}
\label{eq:maxmp}
m \in \argmax_{i=1,2,\ldots,n}  \frac{|\ba_i^T \br_k|}{\|\ba_i\|}.
\end{equation}
The current estimate of $\bx$ is then updated as
\begin{equation}
\bx^{k+1}=\bx^k-\frac{\ba_m^T \br_k}{\|\ba_m\|^2}\be_m,
\end{equation}
and the residual is updated as
\begin{equation}
\br^{k+1}=\bba\bx^{k+1}-\bb=\br_k-\frac{\ba_m^T \br_k}{\|\ba_m\|^2}\ba_m.
\end{equation}
The iterations continue until there are $s$ elements in the support.
Evidently, the MP method coincides with our method as long as the support is smaller than $s$. Our approach however has several advantages:
\begin{itemize}
\item We do not need to initialize it with a zero vector;
\item In MP once an index $m$ is added to the support it will not be removed unless in some iteration $\ba_m^T\br_k=x_m \|\ba_m\|^2$ and $m$ maximizes $\ba_i^T\br_k/\|\ba_i\|$.
 On the other hand, our approach allows to remove elements from the support under much broader conditions. Thus, there is an inherent ``correction" scheme incorporated into our algorithm;
    \item In MP the algorithm stops once the maximal support is achieved. In contrast, in our approach, further iterations are made by utilizing the correction mechanism.
\end{itemize}

We note that once our method converges to a fixed support set, it continues to update the values on the support.
Ultimately, it converges to the least-squares solution on the support since in this situation the method is a simple coordinate descent method employed on a convex function.
 This is similar in spirit to the orthogonal MP (OMP) approach \cite{M08}. The OMP proceeds similarly to the MP method, however, at each stage it updates the vector $\bx^k$ as the least-squares solution on the current support. In our approach, we will converge to the least-squares solution on the final support, however, in choosing the support values we do not perform this orthogonalization.
 Instead, we allow for a correction stage which aids in correcting erroneous decisions.


\subsubsection{Examples}

\begin{example} \label{example:comp} Consider the sparse least squares problem $$ (P_2) \quad \min\{ \|\bba \bx-\bb\|^2 : \bx \in C_2\},$$
 \noindent where $\bba \in \real^{4 \times 5}$  and $\bb \in \real^4$ are given by (up to 4 digits of accuracy):
$$ \begin{pmatrix}   0.8899  & -0.4355  &  0.5304 &  -0.2324  &  0.3745\\
    0.0797  & -0.3475 &   0.0942  &  0.9681 &  -0.4919\\
    0.4425  &  0.3248 &   0.6921  &  0.0921 &   0.7575\\
    0.0773  &  0.7643 &  -0.4804  &  0.0142 &   0.2099
    \end{pmatrix}, \bb = \begin{pmatrix} 1.3254 \\
    0.4272\\
    0.1177\\
   -0.6870
   \end{pmatrix}.$$
The matrix $\bba$ was constructed as follows: first, the components were randomly and independently generated from a standard normal distribution, and then all the columns were normalized. The vector $\bb$ was chosen as $\bb \equiv \bba \bx_{\rm true}$, where $\bx_{\rm true} = (1,-1,0,0,0)^T$, so that $\bx_{\rm true}$ is the optimal solution of the problem. The problem has 10 BF vectors (corresponding to the 5-choose-2 options for the support of the solution) and they are denoted by $1,2,\ldots,10$, where the first solution is the optimal solution $\bx_{\rm true}$. The corresponding objective function values and stationarity levels (with two digits of accuracy) are given in Table \ref{table:funstat}.

 \begin{table}[h]
\begin{center}
\begin{tabular}{||c||c|c|c|c|c|c|c|c|c|c||}\hline\hline
BF vector number & 1 & 2 & 3 & 4 & 5 & 6 & 7 & 8 & 9 & 10\\\hline
function value & -2.42  & -1.60 &  -1.51  & -1.99  & -1.99 &  -1.48 &  -2.11  & -1.33 &  -1.61 &  -0.11\\
stationarity level &     0.00 &    2.90&    8.46&    0.91 &    1.08&   13.97&    0.69 &   18.70 &    1.50 &    9.05\\
\hline \hline
\end{tabular}
\caption{Function values and stationarity levels of the 10 BF vectors of $(P_2)$.}\label{table:funstat}
\end{center}
\end{table}

In this problem $L(f)=4.78$ and $L_2(f)=3.4972$. We compared three methods:
 \begin{itemize} \item the IHT method with $L_1=1.1L(f)$.
 \item the IHT method with $L_2=2L(f)$.
 \item the greedy sparse-simplex method.
 \end{itemize}

Each of these methods was run 1000 times with different randomly generated starting points. All the runs converged to one of the 10 BF vectors. The number of times each method converged to each of the BF vectors is given in Table \ref{table:algconv}.
  \begin{table}[h]
\begin{center}
\begin{tabular}{||c||c|c|c|c|c|c|c|c|c|c||}\hline\hline
BF vector $(i)$ & 1 & 2 & 3 & 4 & 5 & 6 & 7 & 8 & 9 & 10\\\hline
$N_1(i)$ & 329  &   50  &   63  &   92  &  229   &   0 &   130  &   0   & 61  &  46\\
$N_2(i)$ & 340   &  59  &   0   & 89  &    256   &    0  &  187   &   0  &   69  &   0\\
$N_3(i) $& 813    &  0    &  0   & 112  &    0    &  0   &  75   &   0  &    0   &  0\\
$N_4(i)$ & 772 & 0 & 0 & 92 & 0 & 0 & 93 & 0 & 43 & 0 \\
\hline \hline
\end{tabular}
\caption{Distribution of limit points among the 10 BF vector. $N_1(i)$ ($N_2(i)$) is the amount of runs for which the IHT method with $L_1$ ($L_2$) converged to the $i$th BF vector. $N_3(i)$ is the amount of runs for which the greedy sparse-simplex method converged to the $i$th BF vector. The exact definition of $N_4(i)$ will be made clear in Section \ref{sec:maxv}.} \label{table:algconv}
\end{center}
\end{table}

 First note that when employing the IHT method with $L_2=2L(f)=9.56$, the method never converged to the BF vectors 6,8. The theoretical reason for this phenomena is simple: the stationarity levels of these two points are $13.97$ and $18.70$, and they are therefore \textit{not} $9.56$-stationary points. When $L_1=1.1\cdot L(f)=5.26$, there are two additional BF vectors -- 3 and 10 -- to which convergence is impossible, because their stationarity level is $8.46$ and $9.05$. This illustrates the fact that as $L$ gets larger, there are more non-optimal candidates to which the IHT method can converge.
 The greedy sparse-simplex method exhibits the best results with more than $80\%$ chance to converge to the true optimal solution. Note that this method will never converge to the BF vectors  $3,6,8$ and $10$ since they are not $L_2(f)$-stationary points. Moreover, there are only three possible BF vectors  to which the greedy sparse-simplex method converged: $1,4$ and $7$. The reason is that among the 10 BF vectors, there are only three CW-minima. This illustrates the fact that even though any CW-minimum is an $L_2(f)$-stationary point, the reverse claim is not true -- there are $L_2(f)$-stationary points which are not CW-minima.

 In Table \ref{table:singlerun} we describe the 11 first iterations of the greedy sparse-simplex method. Note that at the $4$th iteration the algorithm ``finds" the correct support and the rest of the iterations are devoted to computing the exact values of the nonnegative components of the BF vector.
\begin{table}[h]
\begin{center}
\begin{tabular}{||c||c|c|c|c|c||}\hline\hline
iteration number & $x_1$ & $x_2$ & $x_3$ & $x_4$ & $x_5$\\ \hline
0 & 0 & 1 & 5 & 0 & 0 \\
1 &          0 &    1.0000   &  1.5608  &        0   &       0\\
2 &    0     &    0  &  1.5608    &     0  & -0.6674\\
3 &   1.6431   &      0     &    0      &   0  & -0.6674\\
4 &  1.6431  & -0.8634     &    0        & 0     &    0\\
5 &     1.0290  & -0.8634   &      0     &    0     &    0\\
6 &    1.0290 &  -0.9938      &   0    &     0     &    0\\
7 &    1.0013   &-0.9938     &    0  &       0        & 0\\
8 &    1.0013  & -0.9997      &   0     &    0     &    0\\
9 &    1.0001 &  -0.9997      &   0     &    0    &     0\\
10 &     1.0001  & -1.0000     &    0     &    0     &    0\\
11 &    1.0000  & -1.0000   &      0     &    0    &     0\\
\hline \hline
\end{tabular}
\caption{First 11 iterations of the greedy sparse-simplex method with starting point $(0,1,5,0,0,0)^T$.} \label{table:singlerun}
\end{center}
\end{table}

\end{example}

\begin{example}[Comparison with MP and OMP] To compare the performance of MP and OMP to that of the greedy sparse-simplex, we generated 1000 realizations of $\bba$ and $\bb$
exactly as described in Example \ref{example:comp}. We ran both MP and OMP on these problems with $s=2$. Each of these methods were considered ``successful" if it found
the correct support (MP usually does not find the correct values). The greedy sparse-simplex method was run with an initial vector of all zero, so that the first two iterations
were identical to MP. The results were the following: out of the 1000 realizations both MP and OMP found the correct support in 452 cases.
The greedy sparse-simplex method, which adds  ``correcting" steps to MP was able to recover the correct support in 652 instances.

An additional advantage of the
greedy sparse-simplex method is that it is capable of running from various starting points. We therefore added the following experiment: for each realization of $\bba$ and $\bb$, we ran
the greedy sparse-simplex method from 5 different initial vectors generated in the same way as in Example \ref{example:comp} (and not the all zeros vector). If at least one of these 5 runs detected the correct
support, then the experiment is considered to be a success. In this case the correct support was found 952 times out of the 1000 realizations.

\end{example}

The example above illustrates an important feature of the greedy sparse-simplex algorithm: since it can be initialized with varying starting points, it is possible to improve its performance by using several starting points and obtaining several possible sparse solutions. The final solution can then be taken as the one with minimal objective function value. This feature provides  additional flexibility over the MP and OMP methods.

\subsection{The Partial Sparse-Simplex Method}
\label{sec:maxv}
The greedy sparse-simplex method, as illustrated in Example \ref{example:comp}, has several advantages over the IHT method: first, its limit points satisfy stronger optimality conditions, and as a result is more likely to converge to the optimal solution and second, it does not require knowledge of a Lipschitz constant. On the other hand, the computational effort per iteration of the greedy sparse-simplex algorithm is larger than the one required by the IHT approach. Indeed, in the worst case it requires the call for $O(s \cdot n)$ one-dimensional minimization procedures; this computational burden is caused by the fact that the method has no index selection strategy.
That is, instead of deciding a priori according to some policy on the index or indices on which the optimization will be performed, the algorithm  invokes an optimization procedure for all possible choices, and then picks the index resulting with the minimal objective function value.

The {\it partial sparse-simplex method} described below has an extremely simple way to choose the index or indices on which the optimization will be performed. The only difference from the greedy sparse-simplex algorithm is in the case when   $\|\bx^k\|_0=s$, where  there are two options.
 Either perform a minimization with respect to the variable in the support of $\bx^k$ which causes the maximum decrease in function value; or replace the variable in the support with the smallest absolute value (that is, substituting zero instead of the current value), with the non-support variable corresponding
to the largest absolute value of the partial derivative -- the value of the new non-zero variable is set by performing
a minimization procedure with respect to it. Finally, the best of the two choices (in terms of objective function value) is selected. Since the method is no longer ``greedy" and only considers part of the choices for the pair of indices, we will call it \textit{the partial sparse-simplex method}.

\fcolorbox{black}{Ivory2}{\parbox{15cm}{{\bf The Partial Sparse-Simplex Method}\\

\noindent $\bullet$ {\bf Initialization:} $\bx^0 \in C_s$. \\
\noindent $\bullet$ {\bf General Step ($k=0,1,2,\ldots$):}
\begin{itemize} \item  If $\|\bx^k\|_0<s$, then compute for every $i=1,2,\ldots,n$
\begin{eqnarray*}
 t_i &\in& \argmin_{t \in \real} f(\bx^k+t \be_i),\\
\nonumber f_i &=& \min_{t \in \real} f(\bx^k+t \be_i).
\end{eqnarray*}
Let $i_k \in \argmin_{i=1,\ldots,n} f_i$. If $f_{i_k}<f(\bx^k)$, then set $$\bx^{k+1} = \bx^k+t_{i_k} \be_{i_k}.$$
Otherwise, STOP.
\item If $\|\bx^k\|_0=s$, then compute for every $i \in I_1(\bx^*)$
\begin{eqnarray*}
 t_i &\in& \argmin_{t \in \real} f(\bx^k+t \be_i),\\
\nonumber f_i &=& \min_{t \in \real} f(\bx^k+t \be_i).
\end{eqnarray*}
Let
\begin{eqnarray*} i_k^1 &\in& \argmax \{f_i : i \in I_1(\bx^k)\},\\
i_k^2 & \in& \argmax \{ |\nabla_i f(\bx^k)| : i \in I_0(\bx^k)\},\\
m_k &\in& \argmin \{ |x^k_i| : i \in I_1(\bx^k)\},
\end{eqnarray*}
\noindent and let
$$ \begin{array}{ll} D^1_k = \min_{t \in \real} f(\bx^k+t \be_{i_k^1}), & T^1_k \in \argmin_{t \in \real} f(\bx^k+t \be_{i_k^1})\\
 D^2_k = \min_{t \in \real} f(\bx^k-x_{m_k}^k \be_{m_k}+t \be_{i_k^2}), & T^2_k \in \argmin_{t \in \real} f(\bx^k-x_{m_k}^k \be_{m_k}+t \be_{i_k^2})
\end{array}$$
If $D^1_k<D^2_k$, then set
$$ \bx^{k+1} = \bx^k+T^1_k\be_{i_k^1}.$$
\noindent Else
$$ \bx^{k+1} = \bx^k-x_{m_k}^k \be_{m_k}+T^2_k\be_{i_k^2}.$$

\end{itemize}
}}

\medskip

\begin{remark} The partial sparse-simplex method coincides with the greedy sparse-simplex method when $\|\bx^k\|_0<s$. Therefore, when $f \equiv f_{\rm LI}$, the partial sparse-simplex method coincides with MP for the first $s$ steps and when the initial vector is the vector of all zeros.
\end{remark}
The basic property of the partial sparse-simplex method is that it generates a nonincreasing sequence of function values and that all its limit points are BF vectors.

\begin{lem} \label{lem:pss-bf} Let $\{\bx^k\}$ be the sequence generated by the partial sparse-simplex method. Then any accumulation point of $\{\bx^k\}$ is a BF vector.
\end{lem}
\noindent {\bf Proof.} See Section \ref{sec:lem:pss-bf}.

The limit points of the partial sparse-simplex method are not necessarily CW-minima. However, when Assumption \ref{as:gl} holds, they are $L_2(f)$-stationary points, which is a better result than the one known for the IHT method.
\begin{theorem} \label{thm:ssm} Suppose that Assumption \ref{as:gl} holds and let $\{\bx^k\}$ be the sequence generated by the partial sparse-simplex  method. Then
 any accumulation point of $\{\bx^k\}$ is an $L_2(f)$-stationary point.
\end{theorem}
\noindent {\bf Proof.} See Section \ref{sec:thm:ssm}.

We end this section by returning to Example \ref{example:comp}, and adding a comparison to the partial sparse-simplex algorithm.

\noindent {\bf Example \ref{example:comp} Contd.} In Example \ref{example:comp} we added 1000 runs of the partial sparse-simplex method. The results can be found in Table \ref{table:algconv} under $N_4(i)$, which  is the amount of times in which the algorithm converged to the
$i$th BF vector. As can be seen, the method performs very well, much better than the IHT method with either $L_1=1.1L(f)$ or $L_2=2L(f)$. It is only slightly inferior to the greedy sparse-simplex method since it has another
BS vector to which it might converge (BF vector number 9). Thus, in this example the partial sparse-simplex method is able to compare with the greedy sparse-simplex method despite the fact
that each iteration is much cheaper in terms of computational effort.

\begin{example}[Quadratic Equations] \label{example:quartic} We now consider an example of quadratic equations. Given m vectors
$\ba_1,\ldots,\ba_m$, our problem is to find a vector $\bx \in \real^n$ satisfying:
\begin{eqnarray*}
(\ba_i^T \bx)^2 &=& c_i, \quad i=1,2,\ldots,m,\\
\|\bx\|_0 & \leq & s.
\end{eqnarray*}
The problem can be formulated as problem (P) with $f\equiv f_{\rm QI}$ where $\bba_i = \ba_i \ba_i^T$. We compare the greedy and partial sparse-simplex methods on an example with $m=80,n=120$ and $s=3,4,\ldots,10$. As noted previously in Remark \ref{remark:greedycubic}, the greedy as well as the partial sparse-simplex methods require  to solve several one-dimensional minimization problems of quartic equations at each iteration. Each component of the $80$ vectors $\ba_1,\ldots,\ba_80$ was randomly and independently generated from a standard normal distribution. Then, the ``true" vector $\bx_{\rm true}$ was generated by choosing randomly the $s$ nonzero components whose values were also randomly generated from a standard normal distribution. The vector $\bc$ was then determined by $c_i=(\ba_i^T \bx_{\rm true})^2$. For each value of $s$ ($s=3,4,\ldots,10$), we ran both the greedy and partial sparse-simplex methods from 100 different and randomly generated initial vectors. The numbers of runs out of 100 in which the methods found the correct solution is given in Table \ref{table:quadexample}.
\begin{table}[h]
\begin{center}
\begin{tabular}{||c||c|c||}\hline\hline
s & $N_{\rm PSS}$ & $N_{\rm GSS}$ \\ \hline
3&  27   & 73\\
4&    22  &  69\\
5&     8   & 20\\
6&     5   & 19\\
7 &     9  &  13\\
8&     5   &  8\\
9&      3  &   6\\
10&     2 &    3\\
\hline \hline
\end{tabular}
\caption{The second (third) column contains the number of runs out of 100 for which the partial (greedy) sparse-simplex method converged.} \label{table:quadexample}
\end{center}
\end{table}

As can be clearly seen by the results in the table, the greedy sparse-simplex method outperforms the partial sparse-simplex method in terms of the success probability. In addition, the chances of obtaining the optimal solution decreases as $s$ gets larger. Of course, we can easily increase the success probability of the partial sparse-simplex method by starting it from several initial vectors and taking the best result.
\end{example}
\section{Proofs of Convergence Theorems}
\label{sec:proof}

In this section we collect the main convergence theorems of the algorithms proposed in the previous section.

\subsection{The IHT Method}

\subsubsection{Proof of Theorem~\ref{thm:cong}}
\label{sec:thm:cong}
Suppose that $\bx^*$ is an accumulation point of the sequence. Then there exists a subsequence $\{\bx^{k_n}\}_{n \geq 0}$ that converges to $\bx^*$.
By Lemma \ref{lem:sdescent}
\begin{equation} \label{fxkn}  f(\bx^{k_n})- f(\bx^{k_n+1}) \geq \frac{L-L(f)}{2} \|\bx^{k_n}-\bx^{k_n+1}\|^2.\end{equation}

Since $\{f(\bx^{k_n})\}_{n \geq 0}$ and $\{f(\bx^{k_n+1}) \}_{n \geq 0}$ both converge to the same limit $f^*$, it follows that $f(\bx^{k_n})- f(\bx^{k_n+1}) \rightarrow 0$
as $n \rightarrow \infty$, which combined with (\ref{fxkn}) yields that
$$\bx^{k_n+1} \rightarrow \bx^* \mbox{ as } n \rightarrow \infty.$$
Recall that for all $n \geq 0$
$$ x^{k_n+1} \in P_{C_s} \left ( \bx^{k_n}- \frac{1}{L} \nabla f(\bx^{k_n}) \right ).$$
Let $i \in I_1(\bx^*)$. By the convergence of $\bx^{k_n}$ and $ \bx^{k_n+1}$ to $\bx^*$, it follows that there exists $N$ such that
$$ x_i^{k_n}, x_i^{k_n+1} \neq 0  \mbox{ for all } n>N,$$ and therefore, for $n>N$,
$$ x_i^{k_n+1} = x_i^{k_n}-\nabla_i f(\bx^{k_n}).$$
Taking $n$ to $\infty$ we obtain that
$$ \nabla_i f(\bx^*)=0.$$

Now let $i \in I_0(\bx^*)$. If there exist an infinite number of indices $k_n$ for which $x_i^{k_n+1} \neq 0$, then as in the previous case we obtain
that $ x_i^{k_n+1} = x_i^{k_n}-\nabla_i f(\bx^{k_n})$ for these indices, implying (by taking the limit) that $\nabla_i f(\bx^*)=0$. In particular,
 $\|\nabla_i f(\bx^*)\| \leq L M_s(\bx^*)$.
On the other hand, if there exists an $M>0$ such that for all $n>M$ $x_i^{k_n+1}=0$, then
$$ \left | x_i^{k_n}-\frac{1}{L} \nabla_i f(\bx^{k_n}) \right | \leq M_s\left (\bx^{k_n}-\frac{1}{L}\nabla f(\bx^{k_n}) \right )=M_s(\bx^{k_n+1}).$$
Thus, taking $n$ to infinity while exploiting the continuity of the function $M_s$, we obtain that
$$ \left | \nabla_i f(\bx^*) \right | \leq L  M_s\left (\bx^{*}  \right ),$$
\noindent establishing the desired result. \qed

\subsubsection{Proof of Theorem~\ref{the:conver}}
\label{sec:the:conver}

Let $\{\bx^k\}_{k \geq 0}$ be the sequence generated by the IHT method.  We begin by showing that the sequence is bounded.
By the descent property of the sequence of function values (see Lemma \ref{lem:sdescent}), it follows that the sequence $\{\bx^k\}_{k \geq 0}$ is contained in
the level set
$$ T = \{ \bx \in \real^n : f_{\rm LI} (\bx) \leq f_{\rm LI}(\bx^0)\}.$$

We now show that $T$ is bounded. To this end, note that number of subsets of $\{1,2,\ldots,n\}$ whose size is no larger than $s$ is equal to
$$ p = \sum_{k=0}^s {n \choose k}.$$
By denoting these $p$ subsets as $I_1,I_2,\ldots,I_p$, we can represent the set $T$ as the union:
$$ T = \bigcup_{j=1}^{p} T_j, $$
where

$$ T_j = \left \{ \bx \in \real^n : f_{\rm LI} (\bx) \leq f_{\rm LI} (\bx^0), x_i=0 \quad \forall i \notin I_j \right \}.$$
  In this notation, we can rewrite $T_j$ as
$$ T_j = \left \{ \bx \in \real^n : \|\bba_{T_j} \bx_{T_j} - \bb\|^2 \leq f_{\rm LI}(\bx^0), \bx_{\overline{T_j}} = \bo \right \}.$$
The set $T_j$ is bounded since the $s$-regularity of $\bba$ implies that the matrix $ \bba_{T_j}^T \bba_{T_j}$ is positive definite. This implies the  boundedness of $T$.

We conclude that the sequence $\{\bx^{k}\}_{k \geq 0}$ is bounded and therefore, in particular,
 there exists
a subsequence $\{\bx^{k_n}\}_{n \geq 0}$ which converges to an accumulation point $\bx^*$ which is an $L$-stationary point, and hence also a BF vector. By Lemma \ref{lem:lfinite}, the number of BF vectors is finite, which implies that there exists an $\varepsilon>0$ smaller than the minimal distance between all the pairs of the BF vectors.  To show the convergence of the entire sequence to $\bx^*$, suppose in contradiction that this is not the case. We will assume without loss of generality that the
 subsequence $\{\bx^{k_n}\}_{n \geq 0}$ satisfies $\|\bx^{k_n} -\bx^* \| \leq \varepsilon$ for every $n \geq 0$. Since we assumed that the sequence is not convergent, the index $t_n$
 given by
 $$ t_n = \max \{ l : \|\bx^i-\bx^*\| \leq \varepsilon, i=k_n,k_n+1,\ldots,l\}$$
 \noindent is well defined.
 We have thus constructed a subsequence $\{\bx^{t_n}\}_{n \geq 0}$ for which
 $$ \|\bx^{t_n}-\bx^*\| \leq \varepsilon, \|\bx^{t_n+1}-\bx^*\| > \varepsilon, \quad n=0,1,\ldots$$
It follows that $\bx^{t_n}$ converges to $\bx^*$, and in particular there exists an $N>0$ such that for all $n>N$, $ \|\bx^{t_n}-\bx^*\| \leq \varepsilon/2$.
Thus, for all $n>N$,
$$ \|\bx^{t_n}-\bx^{t_n+1}\| > \frac{\varepsilon}{2},$$
\noindent contradicting Part 3 of Lemma \ref{lem:sdescent}. \qed.

\subsection{The Sparse-Simplex Methods}

\subsubsection{Proof of Theorem~\ref{the:gscw}}
\label{sec:the:gscw}

 By Lemma \ref{lem:bpgls} the sequence of function values $\{f(\bx^k)\}$ is nonincreasing and by Assumption \ref{as:lb} is also bounded below. Therefore, $\{f(\bx^k)\}$ converges.
Suppose that $\bx^*$ is an accumulation point of $\{\bx^k\}$. Then there exists a subsequence $\{\bx^{k_n}\}_{n \geq 0}$ that converges to $\bx^*$. Suppose that $\|\bx^*\|_0=s$. Then the convergence of
$\{\bx^{k_n}\}$ to $\bx^*$ implies that there exists an $N$ such that $I_1(\bx^{k_n}) = I_1(\bx^*)$ for all $n>N$. Let $i \in I_1(\bx^*), j \in \{1,2,\ldots,n\}$ and $t \in \real$.
By definition of the method it follows that
$$ f(\bx^{k_n})-f(\bx^{k_n+1}) \geq f(\bx^{k_n})- f(\bx^{k_n}-x_i^{k_n} \be_i+t \be_j) \mbox{ for all } n >N. $$
The convergence of $\{f(\bx^{k_n})\}$ implies that when taking the limit $n \rightarrow \infty$ in the latter inequality, we obtain
$$ 0 \geq f(\bx^*)-f(\bx^*-x_i^*\be_i +t\be_j).$$
That is, $f(\bx^*) \leq f(\bx^*-x_i^* \be_i+ t \be_j)$ for all $i \in I_1(\bx^*), j\in \{1,2,\ldots,n\}$ and $t \in \real$, meaning that
$$ f(\bx^*) \leq \min_{t \in \real} f(\bx^*-x_i^*\be_i+t \be_j)$$
\noindent for all $i \in I_1(\bx^*)$ and $j \in \{1,2,\ldots,n\}$, thus showing that $\bx^*$ is a CW-minimum. \\

Suppose now that $\|\bx^*\|_0<s$. By the convergence of
$ \{\bx^{k_n}\}$ to $\bx^*$, it follows that there exists an $N$ for which $I_1(\bx^*) \subseteq I_1(\bx^{k_n})$ for all $n>N$. Take $n>N$; if $i \in I_1(\bx^*)$, then $i \in I_1(\bx^{k_n})$, which in particular
implies that
$$ f(\bx^{k_n})-f(\bx^{k_n+1}) \geq f(\bx^{k_n})-f(\bx^{k_n}+t \be_i) \mbox{ for all } t \in \real.$$
Taking $n \rightarrow \infty$ in the last inequality yields the desired inequality
\begin{equation} \label{middle_proof} f(\bx^*) \leq \min_{t \in \real} f(\bx^*+t \be_i).\end{equation}
Now suppose that $i \in I_0(\bx^*)$ and take $n>N$.  If $\|\bx^{k_n}\|_0<s$, then by definition of the greedy sparse-simplex method we have
\begin{equation} \label{d1} f(\bx^{k_n})-f(\bx^{k_n+1}) \geq f(\bx^{k_n})-f(\bx^{k_n}+t\be_i).\end{equation}
On the other hand, if $\|\bx^{k_n}\|_0=s$, then the set $I_1(\bx^{k_n})\setminus I_1(\bx^*)$ is nonempty, and we can pick an index $j_n \in I_1(\bx^{k_n})\setminus I_1(\bx^*)$.
By definition of the greedy sparse-simplex method we have
\begin{equation} \label{d2} f(\bx^{k_n})-f(\bx^{k_n+1}) \geq f(\bx^{k_n})-f(\bx^{k_n}-x_{j_n}^{k_n} \be_{j_n}+t \be_i) \mbox{ for all } t \in \real.\end{equation}
Finally, combining (\ref{d1}) and (\ref{d2}) we arrive at the conclusion that
\begin{equation} \label{d3}  f(\bx^{k_n})-f(\bx^{k_n+1}) \geq f(\bx^{k_n})-f(\bx^{k_n}+\bd_n+t \be_i),\end{equation}
\noindent where
$$\bd_n = \left \{ \begin{array}{ll} 0 & \|\bx^{k_n}\|_0<s,\\ -x_{j_n}^{k_n} \be_{j_n} & \|\bx^{k_n}\|_0=s. \end{array} \right.$$
Since $\bd_n \rightarrow 0$ as $n$ tends to $\infty$, it follows by taking the limit $n \rightarrow \infty$ in (\ref{d3}) the inequality
$$ f(\bx^*) \leq f(\bx^*+t \be_i)$$
\noindent holds for all $t \in \real$, showing that also in this case $\bx^*$ is a CW-minimum. \qed

\subsubsection{Proof of Lemma \ref{lem:pss-bf}}

\label{sec:lem:pss-bf}

The proof of Theorem \ref{the:gscw} until equation (\ref{middle_proof}) is still valid for the partial sparse-simplex method, so that for any $i \in I_1(\bx^*)$
and any $t \in \real$:
$$ f(\bx^*) \leq f(\bx^*+t \be_i),$$
\noindent which in particular means that $ 0 \in \argmin \left \{ g_i(t)\equiv f(\bx^*+t \be_i)\right \}$, and thus $\nabla_i f(\bx^*) =g_i'(0)=0$. \qed

\subsubsection{Proof of Theorem~\ref{thm:ssm}}

\label{sec:thm:ssm}

The proof of the theorem relies on the following lemma:
\begin{lem} \label{lem:useful} Suppose that Assumption \ref{as:gl} holds and let $\{\bx^k\}$ be the sequence generated by the sparse-simplex method. Then for any $k$ for which $\|\bx^k\|_0 <s$ it holds that
\begin{equation} \label{diff1}  f(\bx^k)- f(\bx^{k+1}) \geq \frac{1}{2L_2(f)} \max_{i=1,2,\ldots,n} (\nabla_i f(\bx^k))^2.\end{equation}
For any $k$ with $\|\bx^k\|_0=s$, the inequality
\begin{equation} \label{diff2}  f(\bx^k)-f(\bx^{k+1}) \geq A(\bx^k)\end{equation}
holds true with
\begin{equation} \label{defA} A(\bx) \equiv \max \left \{ \frac{1}{2L_2(f)} \max_{i \in I_1(\bx)} (\nabla_i f(\bx))^2, M_s(\bx)\left [ \max_{i \in I_0(\bx)} |\nabla_i f(\bx)|-\max_{i \in I_1(\bx)} |\nabla_i f(\bx)| - L_2(f)M_s(\bx) \right ] \right \}.
\end{equation}
\end{lem}

\noindent {\bf Proof.} Suppose that $\|\bx^k\|_0 <s$. Then by the definition of the method we have for all $i=1,2,\ldots,n$:
\begin{equation} \label{fxk}  f(\bx^{k+1}) \leq f\left ( \bx^{k}-\frac{1}{L_2(f)}\nabla_{i}f(\bx^{k}) \be_{i} \right ).\end{equation}
\noindent On the other hand, for any $i=1,2,\ldots,n$:
\begin{eqnarray*}  f\left ( \bx^{k}-\frac{1}{L_2(f)}\nabla_{i}f(\bx^{k}) \be_{i} \right ) &\leq& f(\bx^{k})-\frac{1}{L_2(f)} (\nabla_{i}f(\bx^{k}))^2+\frac{1}{2L_2(f)} (\nabla_{i}f(\bx^{k}))^2 \quad (\mbox{Lemma } \ref{lem:localdescent})\\
&=& f(\bx^{k})-\frac{1}{2L_2(f)} (\nabla_{i}f(\bx^{k}))^2,\end{eqnarray*}
\noindent which combined with (\ref{fxk}) implies that
$$ f(\bx^{k})-f(\bx^{k+1}) \geq \frac{1}{2L_2(f)} \max_{i=1,2,\ldots,n}(\nabla_i f(\bx^k))^2,$$
\noindent establishing (\ref{diff1}).

Next, suppose that $\|\bx^k\|_0=s$. A similar argument to the one just invoked shows that
 \begin{equation} \label{l2x} f(\bx^{k})-f(\bx^{k+1}) \geq \frac{1}{2L_2(f)} \max_{i \in I_1(\bx^k)}(\nabla_i f(\bx^k))^2.\end{equation}
By the definition of the greedy sparse-simplex method, it follows that
\begin{equation} \label{994}  f(\bx^{k})-f(\bx^{k+1}) \geq f(\bx^{k}) -f(\bx^{k}-x_{m_k}^k\be_{m_k} +T^2_{k}\be_{i_{k}^2})
\geq f(\bx^{k})-f(\bx^{k}-x_{m_k}^{k} \be_{m_k}-\sigma x_{m_k}^{k}\be_{i_{k}^2}),\end{equation}
\noindent where $\sigma=\sgn(x_{m_k}^{k} \nabla_{i_{k}^2} f(\bx^{k}))$. Using the local descent lemma (Lemma \ref{lem:localdescent}) once more, we obtain that
\begin{eqnarray} \nonumber && f(\bx^{k} - x_{m_k}^{k} \be_{m_k} - \sigma x_{m_k}^{k} \be_{i_{k}^2}) \\
\nonumber &&\leq f(\bx^{k}) +\nabla f(\bx^{k})^T (-x_{m_{k}}^{k} \be_{m_k} - \sigma x_{m_k}^k \be_{i_{k}^2}) +\frac{L_2(f)}{2} \left \|-x_{m_k}^{k} \be_{m_k} - \sigma x_{m_k}^k \be_{i_{k}^2} \right \|^2 \\
\nonumber && = f(\bx^{k}) -x_{m_k}^{k} \nabla_{m_k} f(\bx^{k}) -\sigma x_{m_k}^{k} \nabla_{i_{k}^2} f(\bx^{k})+L_2(f) (x_{m_k}^{k})^2\\
&& = f(\bx^{k}) + M_s(\bx^{k}) \left [L_2(f)M_s(\bx^{k})-|\nabla_{i_{k}^2}f(\bx^k)|\right ]-x_{m_k}^k \nabla_{m_k} f(\bx^k). \label{995}
\end{eqnarray}
Combining (\ref{994}) and (\ref{995}) we obtain that
\begin{equation} \label{996} f(\bx^k)-f(\bx^{k+1}) \geq M_s(\bx^k) \left [ \max_{i \in I_0(\bx^k)} | \nabla_i f(\bx^k)| -L_2(f) M_s(\bx^k) \right ]+x_{m_k}^k \nabla_{m_k} f(\bx^k). \end{equation}
Finally, (\ref{l2x}) and (\ref{996}) along with the fact that
$$ x_{m_k}^k \nabla_{m_k} f(\bx^k) \geq -M_s(\bx^k) \max_{i \in I_1(\bx^k)} |\nabla_i f(\bx^k)|$$
\noindent readily imply the inequality (\ref{diff2}). \qed

We now turn to prove Theorem~\ref{thm:ssm}.
   Let $\bx^*$ be an accumulation point of the generated sequence. Then there exists a subsequence $\{\bx^{k_n}\}_{n \geq 0}$ converging to $\bx^*$.
Suppose first that $\|\bx^*\|_0=s$. Then there exists an $N>0$ such that $I_1(\bx^{k_n}) = I_1(\bx^*)$ for all $n>N$. Therefore, by (\ref{diff2}) we have
\begin{equation} \label{ineq1} f(\bx^{k_n})-f(\bx^{k_n+1}) \geq A(\bx^{k_n})\end{equation}
\noindent for all $n>N$.  Since $\{f(\bx^k)\}$ is a nonincreasing and lower bounded sequence, it follows that the left hand side of the inequality (\ref{ineq1}) tends to $0$ as $n \rightarrow \infty$.
Therefore, by the continuity of the operator $A$ we have $A(\bx^*)\leq 0$ from which it follows that
\begin{eqnarray}
 \label{1018} \frac{1}{2L_2(f)} \max_{i \in I_1(\bx^*)} (\nabla_i f(\bx^*))^2 &=&0, \\
 \label{1019}  M_s(\bx^*)\left [ \max_{i \in I_0(\bx^*)} |\nabla_i f(\bx^*)|-\max_{i \in I_1(\bx^*)}
 |\nabla_i f(\bx^*)| - L_2(f)M_s(\bx^*) \right ] &\leq&0.
\end{eqnarray}

By (\ref{1018}) it follows that $\nabla_i f(\bx^*)=0$ for all $i \in I_1(\bx^*)$ and substituting this in (\ref{1019}) yields the inequality
$$  \max_{i \in I_0(\bx^*)} |\nabla_i f(\bx^*)| \leq L_2(f) M_s(\bx^*),$$
\noindent meaning that $\bx^*$ is an $L_2(f)$-stationary point. \\
Now suppose that $\|\bx^*\|_0<s$. There are two cases. If there exists an infinite number of $n$-s for which $\|\bx^{k_n}\|_0<s$,
then by Lemma \ref{lem:useful} for each such $n$
$$ f(\bx^{k_n})-f(\bx^{k_n+1}) \geq \frac{1}{2L_2(f)} \max_{i=1,2,\ldots,n} \nabla_i f(\bx^{k_n})^2,$$
and therefore by taking $n \rightarrow \infty$ along the $n$-s for which $\|\bx^{k_n}\|_0<s$, we obtain that $\nabla f(\bx^*)=0$.
If, on the other hand, there exists an integer $N$ such that the equality $\|\bx^{k_n}\|_0=s$  holds for all $n>N$,
then by the definition of the method we have for all $n>N$
\begin{equation} \label{2378} f(\bx^{k_n})-f(\bx^{k_n+1}) \geq \frac{1}{2L_2(f)} \max_{i \in I_1(\bx^{k_n})} (\nabla_i f(\bx^{k_n}) )^2\end{equation}
\noindent and
\begin{eqnarray} \label{1031} f(\bx^{k_n})-f(\bx^{k_n+1}) &\geq& f(\bx^{k_n})-f(\bx^{k_n}-x_{m_k}^k\be_{m_k}+T_k^2\be_{i_k^2})\\
\nonumber &=& f(\bx^{k_n})-f(\bx^{k_n}-x_{m_k}^k \be_{m_k})+f(\bx^{k_n}-x_{m_k}^k \be_{m_k})-f(\bx^{k_n}-x_{m_k}^k\be_{m_k}+T_k^2\be_{i_k^2}).
\end{eqnarray}
Since $T_k^2 \in \argmin_{t \in \real} f(\bx^{k_n}-x_{m_k}^k\be_{m_k}+t\be_{i_k^2})$, then
\begin{eqnarray*}  f(\bx^{k_n}-x_{m_k}^k \be_{m_k})-f(\bx^{k_n}-x_{m_k}^k\be_{m_k}+T_k^2\be_{i_k^2}) &\geq& \frac{1}{2L_2(f)} (\nabla_{i_k^2} f(\bx^{k_n}-x_{m_k}^k \be_{m_k}))^2\\
&=&\frac{1}{2L_2(f)} \max_{i \in I_0(\bx^k)}(\nabla_{i} f(\bx^{k_n}-x_{m_k}^k \be_{m_k}))^2,
\end{eqnarray*}
\noindent which combined with (\ref{1031}) yields
\begin{equation} \label{1042} \frac{1}{2L_2(f)} \max_{i \in I_0(\bx^k)}(\nabla_{i} f(\bx^{k_n}-x_{m_k}^k \be_{m_k}))^2\leq f(\bx^{k_n}-x_{m_k}^k \be_{m_k})-f(\bx^{k_n+1}).\end{equation}
In addition,
\begin{eqnarray*} |\nabla_i f(\bx^{k_n})| &\leq& |\nabla_i f(\bx^{k_n})-\nabla_i f(\bx^{k_n}-x_{m_k}^k \be_{m_k})|+|\nabla_i f(\bx^{k_n}-x_{m_k}^k \be_{m_k})|\\
&\leq& L_2(f) |x_{m_k}^k| +|\nabla_i f(\bx^{k_n}-x_{m_k}^k \be_{m_k})| \\
&=& L_2(f)M_s(\bx^{k_n}) +|\nabla_i f(\bx^{k_n}-x_{m_k}^k \be_{m_k})|,\end{eqnarray*}
\noindent and thus (\ref{1042}) readily implies that:
$$  \max_{i \in I_0(\bx^k)}|\nabla_{i} f(\bx^{k_n})|\leq L_2(f)M_s(\bx^{k_n}) +\sqrt{2L_2(f) [f(\bx^{k_n}-x_{m_k}^k \be_{m_k})-f(\bx^{k_n+1})]},$$
\noindent which together with (\ref{2378}) yields that for all $i=1,2,\ldots,n$
{\small
$$ |\nabla_{i} f(\bx^{k_n})| \leq \min \left \{ L_2(f)M_s(\bx^{k_n}) +\sqrt{2L_2(f) [f(\bx^{k_n}-x_{m_k}^k \be_{m_k})-f(\bx^{k_n+1})]}, \sqrt{2L_2(f)f(\bx^{k_n})-f(\bx^{k_n+1})}\right \}.$$
}
Since the righthand side of the latter inequality converges to 0 as $n \rightarrow \infty$, it follows that the desired result $\nabla f(\bx^*)=0$ holds. \qed

\bibliographystyle{plain}
\bibliography{notes}

\begin{thebibliography}{10}

\bibitem{BT09-book}
A.~Beck and M~Teboulle.
\newblock Gradient-based algorithms with applications to signal recovery
  problems.
\newblock In Yonina Eldar and Daniel Palomar, editors, {\em Convex Optimization
  in Signal Processing and Communications}. Cambridge University Press, 2010.

\bibitem{BF12}
E.~V.~D. Berg and M.~P. Friedlander.
\newblock Sparse optimization with least-squares constraints.
\newblock {\em SIAM J. Optim.}, 21:1201--1229.

\bibitem{B99}
D.~P. Bertsekas.
\newblock {\em Nonlinear Programming}.
\newblock Belmont MA: Athena Scientific, second edition, 1999.

\bibitem{BD08}
T.~Blumensath and M.~E. Davies.
\newblock Iterative thresholding for sparse approximations.
\newblock {\em The Journal of Fourier Analysis and Applications},
  14(5):629--654, 2008.

\bibitem{BD10}
T.~Blumensath and M.~E. Davies.
\newblock Normalised iterative hard thresholding; guaranteed stability and
  performance.
\newblock {\em IEEE Journal of Selected Topics in Signal Processing},
  4:298--309, 2010.

\bibitem{CRT06}
E.~Cand\`{e}s, J.~Romberg, and T.~Tao.
\newblock Robust uncertainty principles: {E}xact signal reconstruction from
  highly incomplete frequency information.
\newblock {\em IEEE Trans. Inform. Theory}, 52(2):489--509, 2006.

\bibitem{D98}
R.~DeVore.
\newblock Nonlinear approximation.
\newblock {\em Acta Numerica}, 7:51--150, 1998.

\bibitem{D95}
D.~Donoho.
\newblock Denoising by soft-thresholding.
\newblock {\em IEEE Trans. Inform. Theory}, 41(3):613--627, 1995.

\bibitem{D06}
D.~L. Donoho.
\newblock Compressed sensing.
\newblock {\em IEEE Transactions on Information Theory}, 52:1289--1306, 2006.

\bibitem{DE03}
D.~L. Donoho and M.~Elad.
\newblock Optimally sparse representation in general (non-orthogonal)
  dictionaries via l1 minimization.
\newblock In {\em PROC. of the National Academy of Sciences}, volume 100, pages
  2197--2202, 2003.

\bibitem{SEA12}
A.~Szameit et. al.
\newblock Sparsity-based single-shot sub-wavelength coherent diffractive
  imaging.
\newblock {\em Nature Materials}.

\bibitem{F82}
J.~R. Fienup.
\newblock Phase retrieval algorithms: a comparison.
\newblock {\em Applied Optics}, 21:2758--2769, 1982.

\bibitem{GS72}
R.~W. Gerchberg and W.~O. Saxton.
\newblock A practical algorithm for the determination of phase from image and
  diffraction plane pictures.
\newblock {\em Optik}, 35:237--246, 1972.

\bibitem{GR97}
I.~F. Gorodnitsky and B.~D. Rao.
\newblock {Sparse signal reconstruction from limited data using FOCUSS: A
  re-weighted minimum norm algorithm}.
\newblock {\em IEEE Trans. Signal Processing}, 45(3):600--616, Mar. 1997.

\bibitem{H89}
N.~Hurt.
\newblock {\em Phase Retrieval and Zero Crossings}.
\newblock Norwell, MA: Kluwer Academic Publishers, 1989.

\bibitem{DDEK11}
Y.~C.~Eldar M.~Davenport, M.~Duarte and G.~Kutyniok.
\newblock {\em Compressed Sensing: Theory and Applications}, chapter
  Introduction to Compressed Sensing.
\newblock Cambridge Univ. Press, 2012.

\bibitem{M08}
S.~Mallat.
\newblock {\em {A Wavelet Tour of Signal Processing: The Sparse Way}}.
\newblock Academic Press, 2008.

\bibitem{MZ93}
S.~Mallat and Z.~Zhang.
\newblock Matching pursuits with time-frequency dictionaries.
\newblock {\em IEEE Trans. Signal Processing}, 41(12):3397--3415, 1993.

\bibitem{OF96}
B.~Olshausen and D.~Field.
\newblock Emergence of simple-cell receptive field properties by learning a
  sparse representation.
\newblock {\em Nature}, 381:607--609, 1996.

\bibitem{SESS11}
Y.~Shechtman, Y.~C. Eldar, A.~Szameit, and M.~Segev.
\newblock Sparsity-based sub-wavelength imaging with partially spatially
  incoherent light via quadratic compressed sensing.
\newblock {\em Optics Express}, 19:14807--14822, 2011.

\bibitem{JPEG2000}
D.~Taubman and M.~Marcellin.
\newblock {\em {JPEG} 2000: {I}mage Compression Fundamentals, Standards and
  Practice}.
\newblock Kluwer, 2001.

\bibitem{T96}
R.~Tibshirani.
\newblock Regression shrinkage and selection via the lasso.
\newblock {\em J. Royal Statist. Soc B}, 58(1):267--288, 1996.

\bibitem{T04}
J.~Tropp.
\newblock Greed is good: {A}lgorithmic results for sparse approximation.
\newblock {\em IEEE Trans. Inform. Theory}, 50(10):2231--2242, October 2004.

\bibitem{TW10}
J.~Tropp and S.~J. Wright.
\newblock Computational methods for sparse solution of linear inverse problems.
\newblock {\em Proc. IEEE}, 98(6):948--958, 2010.

\bibitem{V11}
R.~Vershynin.
\newblock {\em Compressed Sensing: Theory and Applications}, chapter
  Introduction to the non-asymptotic analysis of random matrices.
\newblock Cambridge Univ. Press, 2012.

\end{thebibliography}

\end{document}